\documentclass[12pt,twoside,a4paper]{article}
\pdfoutput=1
\usepackage{graphicx}
\usepackage{url}
\usepackage{color}
\usepackage{cite}

\def\gsim{\:\raisebox{-0.5ex}{$\stackrel{\textstyle>}{\sim}$}\:}
\begin{document}
\thispagestyle{empty} 
\title{
\vskip-3cm
{\baselineskip14pt
\centerline{\normalsize DESY 20-058 \hfill ISSN 0418--9833}
\centerline{\normalsize MITP/20--015 \hfill} 
\centerline{\normalsize April 2020 \hfill}} 
\vskip1.5cm
\boldmath
{\bf $\Lambda_c^{\pm}$ production in $pp$ collisions}
\\
{\bf with a new fragmentation function}
\unboldmath
\author{
B.~A.~Kniehl$^1$, 
G.~Kramer$^1$, 
I.~Schienbein$^2$ 
and H.~Spiesberger$^3$
\vspace{2mm} \\
\normalsize{
  $^1$ II. Institut f\"ur Theoretische
  Physik, Universit\"at Hamburg,
}\\ 
\normalsize{
  Luruper Chaussee 149, D-22761 Hamburg, Germany
} \vspace{2mm}\\
\normalsize{
  $^2$ Laboratoire de Physique Subatomique et de Cosmologie,
} \\ 
\normalsize{
  Universit\'e Joseph Fourier Grenoble 1,
}\\
\normalsize{
  CNRS/IN2P3, Institut National Polytechnique de Grenoble,
}\\
\normalsize{
  53 avenue des Martyrs, F-38026 Grenoble, France
} \vspace{2mm}\\
\normalsize{
  $^3$ PRISMA$^+$ Cluster of Excellence, Institut f\"ur Physik,
}\\ 
\normalsize{
  Johannes-Gutenberg-Universit\"at,
  Staudinger Weg 7, D-55099 Mainz, Germany}\\
\vspace{2mm} \\}}
\maketitle

\begin{abstract}
\medskip
\noindent
We study inclusive $\Lambda_c^{\pm}$-baryon production in $pp$ 
collisions in the general-mass variable-flavor number scheme and 
compare with data from the LHCb, ALICE and CMS collaborations. 
We perform a new fit of the $c \to \Lambda_c^+$ fragmentation 
function combining $e^+e^-$ data from OPAL and Belle. The 
agreement with LHC data is slightly worse compared with a 
calculation using an older fragmentation function, and the 
tension between different determinations of $\Lambda_c^{\pm}$ 
production cross sections from the LHC experimental 
collaborations is not resolved. The ratio of data for 
$\Lambda_c^+$-baryon and $D^0$-meson production seems to 
violate the universality of $c$-charm quark to $c$-hadron 
fragmentation. 
\\
\\
PACS: 12.38.Bx, 12.39.St, 13.85.Ni, 14.40.Nd
\end{abstract}

\clearpage


\section{Introduction}

The inclusive production of hadrons containing the heavy 
charm or bottom quarks, $c$ and $b$, plays an important 
role in testing quantum chromodynamics (QCD). The predictions 
in the framework of perturbative QCD are based on the 
factorization approach. Cross sections are calculated as 
a convolution of three terms: the parton distribution 
functions (PDFs) describing the parton content of the 
initial hadronic state (for a review, see 
Ref.~\cite{Tanabashi:2018oca}), the partonic hard-scattering 
cross sections calculated as perturbative series in 
powers of the strong-coupling constant, and the fragmentation 
functions (FFs), which describe the production yield and the 
momentum distribution for a given heavy-quark hadron 
originating from a parton. The PDFs and the FFs are 
non-perturbative objects and determined from experimental 
data. 

Inclusive production of charmed baryons, in particular of 
$\Lambda_c^{\pm}$, is of interest for several reasons. First, 
there is the general question whether the perturbative 
approach to calculate production cross sections applies 
to $c$-baryons in the same way as it does for $c$-mesons. 
The details of the fragmentation mechanism of $c$-quarks 
and other partons, for example gluons, into $c$-baryons 
and into $c$-mesons may be different, in particular at 
small transverse momentum. The new data from the 
experiments at the LHC are expected to provide us with 
valuable information to answer this question. Actually, 
there are indications that, in the case of bottom quarks, 
universality of the $\Lambda_b^0$ fragmentation process is 
violated as discussed in detail in Ref.~\cite{Amhis:2016xyh}. 
It was shown that the $b \to \Lambda_b^0$ fragmentation fraction 
obtained from $pp$ and $p\bar{p}$ data is not compatible 
with that deduced from LEP data. Similar discrepancies might 
be expected for the case of charmed-baryon production.

So far, FFs for charmed hadrons have been derived from data 
obtained in $e^+e^-$ annihilation. These data are well suited 
for a determination of FFs since they are free from uncertainties 
due to the properties of the initial state. Very detailed and 
precise data for charmed-meson and -baryon production, including 
$D^0$, $\bar{D^0}$, $D^{\pm}$, $D^{*\pm}$, $D_s^{\pm}$, and 
$\Lambda_c^{\pm}$, have been obtained by the Belle collaboration 
at the KEK storage ring \cite{Seuster:2005tr}. Fragmentation 
functions for the $D$ mesons have been obtained from these 
data combined with earlier LEP data from the OPAL collaboration 
by T.\ Kneesch and three of us \cite{Kneesch:2007ey}. A fit 
of the $c \to \Lambda_c^+$ FF was, however, not performed in 
that work. 

A common ansatz for heavy-quark FFs was given by Peterson 
et al.~\cite{Peterson:1982ak} and is defined by
\begin{equation}
D_c(x,\mu_0) = N \frac{x(1-x)^2}{[(1-x)^2+\epsilon x]^2} 
\, , 
\end{equation} 
where $x$ is the momentum fraction transferred from the 
charm quark to the observed charm hadron and $N$ and 
$\epsilon$ are parameters fitted to data at an initial 
scale $\mu_0$. The FF is evolved to larger scales $\mu > 
\mu_0$ by the DGLAP evolution equations. Two of us have used 
this ansatz in 2005 and 2006 \cite{Kniehl:2005de,Kniehl:2006mw} 
to obtain the first FFs for $\Lambda_c^{\pm}$. The FFs in 
these two references differed in the initial condition. 
The starting scale in Ref.~\cite{Kniehl:2005de} was $\mu_0 
= 2m_c$, and in Ref.~\cite{Kniehl:2006mw} $\mu_0 = m_c$ was 
chosen instead, with the charm quark mass $m_c = 1.5$~GeV. 
The FFs were obtained from fits to data of the normalized 
differential cross section $(1/\sigma_{tot}) d\sigma/dx$ for 
$e^+e^- \to \gamma /Z \to \Lambda_c^{\pm}$ measured by the 
OPAL collaboration at LEP1 \cite{Alexander:1996wy}. The OPAL 
data include $b$-tagged events where the $\Lambda_c^{\pm}$ 
baryons originate from $b$-quarks. This contribution was 
fitted in Ref.~\cite{Kniehl:2005de,Kniehl:2006mw} with a 
power ansatz: 
\begin{equation}
D_b(x,\mu_0^b) =N x^{\alpha}(1-x)^{\beta} 
\, , 
\end{equation} 
with $\mu_0^b = 5$~GeV. The parameter values for $N$, 
$\alpha$, $\beta$ and $\epsilon$ can be found in 
Refs.~\cite{Kniehl:2005de,Kniehl:2006mw}. 

The $\Lambda_c^+$ FFs constructed in Ref.~\cite{Kniehl:2006mw} 
were later used to predict cross sections in the 
general-mass variable-flavour-number scheme (GM-VFNS) 
for inclusive $\Lambda_c^{\pm}$ production at the LHC, see 
Ref.~\cite{Kniehl:2012ti}, where the center-of-mass 
energy ($\sqrt{S} = 7$~TeV) is much higher than the 
energy at which the FFs have been fitted to data 
($\sqrt{S} = M_Z$). Corresponding measurements of 
inclusive $\Lambda_c^{\pm}$ production in $pp$ collisions at 
LHC energies have been done first by the LHCb collaboration 
\cite{Aaij:2013mga}. These measurements are in reasonably 
good agreement with the previous calculations of 
Ref.~\cite{Kniehl:2012ti} at transverse momenta $2 < p_T < 
8$~GeV and for the rapidity range $2.0 < y < 4.5$. The more 
recent ALICE data for inclusive $\Lambda_c^{\pm}$ production 
at $\sqrt{S} = 7$~TeV in the central rapidity region $|y| < 
0.5$ \cite{Acharya:2017kfy} have been compared to a calculation 
with the FFs from Ref.~\cite{Kniehl:2006mw} in the range 
$3 < p_T <8$~GeV. In this case, the predictions underestimate 
the ALICE data \cite{Acharya:2017kfy}: the latter have been 
found to be larger by a factor of 2.5, on average, than the 
result of the GM-VFNS calculation at the default choice for 
factorization and renormalization scales, and outside the 
theoretical uncertainties obtained from variations of the 
scale parameters. The most recent measurements of 
$d\sigma/dp_T$ for inclusive $\Lambda_c^{\pm}$ production are 
from the CMS collaboration at the LHC \cite{Sirunyan:2019fnc}. 
These measurements have been done at $\sqrt{S}=5.02$~TeV in the 
rapidity interval $|y| < 1.0$ for $p_T$ between $5$ and 
$20$~GeV. We will compare these data with GM-VFNS calculations 
using the FFs from Ref.~\cite{Kniehl:2006mw} below. 

A good measure of the strength of inclusive $\Lambda_c^{\pm}$ 
production for various scattering processes and kinematical 
conditions is the $\Lambda_c^+/D^0$ production ratio. A 
detailed knowledge of this ratio, including its dependence on 
kinematic variables like $p_T$ and $y$, should be sensitive 
to the fragmentation mechanism in the charm sector. 
Related important quantities characterizing the strength 
of the hadronization of a heavy quark into a hadron  
are the so-called fragmentation fractions defined by 
\begin{equation}
B_Q(\mu) 
= 
\int_{x_{cut}}^{1}dxD_Q(x,\mu) 
\, , 
\end{equation}
and the average energy fractions
\begin{equation}
x_Q(\mu) 
= 
\frac{1}{B_Q(\mu)} \int_{x_{cut}}^{1}dx \, x D_Q(x,\mu) 
\, . 
\end{equation} 
For the FFs constructed in Ref.~\cite{Kniehl:2006mw}, these 
fragmentation fractions in the next-to-leading order (NLO) 
fit turned out to be $B_c(2m_c) = 0.0612$ and $B_b(2m_b) = 
0.143$. The corresponding average momentum fractions for 
$c,b \to \Lambda_c^+$ were found to be $x_c(2m_c) = 0.738$ 
and $x_b(2m_b) = 0.290$. These are the values at threshold. 
Corresponding values at $\mu = M_Z$ can be found in 
Ref.~\cite{Kniehl:2006mw}.

In principle, these quantities can be determined from 
integrated cross sections. However, data are usually available 
in a restricted phase space region only and an extrapolation 
to the full phase space is needed. In addition, experimental 
determinations are averaged over phase space in such a way 
that the (weak) scale dependence of $B_Q$ and $x_Q$ is lost. 

Recently, a summary of experimental data for the fragmentation 
fractions of charm quarks into specific charmed hadrons was 
given in Ref.~\cite{Lisovyi:2015uqa}. Measurements performed 
in photoproduction, deep inelastic $e^{\pm}p$ scattering, 
$pp$ collisions and $e^+e^-$ annihilation were compared 
on the basis of up-to-date branching ratios for the respective 
decays of the final charmed hadrons. In that work, the average 
branching fraction for the transition $c \to \Lambda_c^+$ 
in $Z$ decays was found to be $B_c=0.060 \pm 0.0177$. The 
value from Ref.~\cite{Lisovyi:2015uqa} is based not only on OPAL 
measurements \cite{Alexander:1996wy}, but data from ALEPH and 
DELPHI were included as well. The branching fraction determined 
from other processes came out quite similar: 
$B_c = 0.0540\pm 0.0195$ (from $e^{\pm}p$ DIS data), 
$B_c = 0.067 \pm 0.0106$ (from photoproduction in $e^{\pm}p$ 
scattering) and $B_c = 0.0639 \pm 0.0122$ (from $pp$ 
collisions at the LHC). Also $\Lambda_c^{\pm}$ production 
in $e^+e^-$ annihilation at a center-of-mass energy 
$\sqrt{S} = 10.5$~GeV was analysed. The results are based on 
measurements performed by CLEO \cite{Bortoletto:1988kw, 
Avery:1990bc}, ARGUS \cite{Albrecht:1988an}, BABAR 
\cite{Aubert:2006cp} and Belle \cite{Seuster:2005tr}. 
From these data, the branching fraction $B_c = 0.0611 \pm 
0.0060$ was obtained. This value agrees with the one obtained 
directly by BABAR  \cite{Aubert:2006cp}, i.e.\ $B_c = 0.071 
\pm0.003 \pm 0.018$, where the second error is due to the 
uncertainty of the branching fraction for the reconstructed 
decay mode. From the analysis of Ref.~\cite{Lisovyi:2015uqa}, 
we can conclude that all branching fractions agree very well 
between the different production channels. This is in 
contrast with corresponding observations in the $b$-quark 
sector, i.e.\ for fragmentation fractions of $b \to 
\Lambda_b^0$ transitions \cite{Kramer:2018rgb}. There, data 
from the LEP and Tevatron experiments disagree and indicate 
a strong dependence on kinematic properties of the production 
process. 

All these determinations are in good agreement with the 
value $B_c(2m_c)=0.0612$ obtained in the fit reported in 
Ref.~\cite{Kniehl:2006mw}. However, we have to note that 
the value in Ref.~\cite{Kniehl:2006mw} is connected with an 
obsolete value for the branching ratio Br$(\Lambda_c^+ \to 
\pi^+K^-p) = 0.044$ as it was known in 1996, while the 
recent PDG value is Br$(\Lambda_c^+ \to \pi^+K^-p) = 0.0635$ 
\cite{Tanabashi:2018oca}. The agreement is therefore fortuitous. 

In the following, we present predictions for the production 
of $\Lambda_c^{\pm}$ in the LHC experiments, first using the 
FF fit based only on the high-energy LEP data from OPAL 
\cite{Alexander:1996wy}. Then, a new fit where we include also 
$e^+e^- \to \Lambda_c^{\pm} +X $ data at $\sqrt{S} = 10.5$~GeV 
from Belle \cite{Niiyama:2017wpp} and corresponding 
predictions for the LHC will be presented.

The outline of the paper is as follows. In Sect.~2, we 
introduce our strategy for the calculation of cross sections 
for inclusive $\Lambda_c^{\pm}$ production in the LHCb, ALICE and 
CMS experiments. In Sect.~3, we present numerical results for 
the $p_T$-differential cross sections and compare them with these 
three LHC experiments. These calculations are performed using the 
old FF of Ref.~\cite{Kniehl:2006mw} to describe the transition 
of $c$ and $b$ quarks to $\Lambda_c^+$. A new FF set obtained 
from a fit to the OPAL data \cite{Alexander:1996wy} and 
the new measurements from Belle at $\sqrt{S}=10.52$~GeV 
\cite{Niiyama:2017wpp} are then described in Sect.~4. This 
new FF is used in Sect.~5 for a calculation of $d\sigma/dp_T$ 
for $\Lambda_c^{\pm}$ production and compared with the LHC data. 
A discussion of our findings and conclusions are presented 
in Sect.~5.

\section{Setup and input 
\label{Sec:input}
}

The theoretical description of the GM-VFNS framework as 
well as technical details of its implementation can be 
found in Refs.~\cite{Kniehl:2004fy,Kniehl:2005mk}. Here 
we describe only the input required for the numerical 
computations for which the results are given in the next 
section. For the proton PDF, we use CTEQ14 \cite{Dulat:2015mca} 
as implemented in the LHAPDF library \cite{Buckley:2014ana}. 
We take the $c$-quark pole mass to be $m_c=1.3$~GeV, in 
agreement with the value assumed for the PDF set CTEQ14. 
The strong coupling $\alpha_s^{(n_f)}(\mu_R)$ is evaluated 
at NLO with $\Lambda_{\overline{\rm MS}}^{(4)}=328$~MeV. This 
corresponds to $\Lambda_{\overline{\rm MS}}^{(5)} = 225$~MeV 
above the 5-flavour threshold chosen at $m_b = 5$~GeV.

In the following sections, we shall take equal values for 
the initial- and final-state factorization scales $\mu_F$, 
entering the PDFs and FFs, respectively. We choose $\mu_F$ 
and the renormalization scale $\mu_R$ at which $\alpha_s$ 
is evaluated as 
\begin{equation}
\mu_F = 0.98 \mu_T \, , \quad 
\mu_R = \xi_R \mu_T \, , 
\label{eq:scales}
\end{equation}
with $\mu_T = 0.5 \sqrt{p_T^2+4m_c^2}$ and $m_c = 1.3$~GeV. 
Theoretical uncertainties will be estimated by varying $\xi_R$ 
in the range between $1/2$ and $2$. This choice, in particular 
for $\mu_F$, allows us to obtain realistic predictions for 
$p_T$ values also below 3~GeV, as needed to compare with the 
complete data set of the LHCb collaboration as well as with 
ALICE data. The choice in Eq.~(\ref{eq:scales}) for $\mu_F$ 
was first applied in Ref.~\cite{Benzke:2017yjn}, where we 
found it instrumental to obtain good agreement with inclusive 
$D$-meson cross sections $d\sigma/dp_T$ for $p_T$ values down 
to $p_T = 0$.

First, we shall use the non-perturbative $\Lambda_c^+$ FFs 
constructed in our earlier work \cite{Kniehl:2006mw}. These 
FFs have been fitted only to inclusive $(\Lambda_c^+ + c.c.)$ 
production data in $e^+e^-$ annihilation on the $Z$ resonance 
measured by the OPAL collaboration at LEP1 \cite{Alexander:1996wy}. 
The starting scale was fixed at $\mu_0 = 1.5$~GeV. In a subsequent 
section, we shall describe a new $\Lambda_c^+$ FF obtained from 
a fit to data which include, in addition to the old OPAL data, 
$e^+e^-$ data at $\sqrt{S} = 10.52$~GeV from the Belle 
collaboration~\cite{Niiyama:2017wpp}.

\section{Comparison with LHCb, ALICE and CMS data using old FFs}

The first measurements of the cross section for inclusive 
$\Lambda_c^{\pm}$ production in $pp$ collisions at the LHC were 
performed some time ago by the LHCb collaboration 
\cite{Aaij:2013mga}. These measurements provided data for 
differential cross sections $d\sigma/dp_T$ for $\Lambda_c^+ 
+ c.c.$ baryons in bins of $p_T$, integrated over the forward 
rapidity range $2.0 < y < 4.5$. In addition, cross sections 
in bins of $y$ integrated over the $p_T$ range $2 < p_T < 8$~GeV 
were presented. The $p_T$-bin integrated cross sections 
were compared with our GM-VFNS calculations using the 
$\Lambda_c^+$ FFs of Ref.~\cite{Kniehl:2006mw} for 
$p_T \geq 3$~GeV. These GM-VFNS predictions agreed 
fairly well with the LHCb data \cite{Aaij:2013mga}. 

\begin{figure*}[b!]
\begin{center}
\includegraphics[height=0.37\textheight]{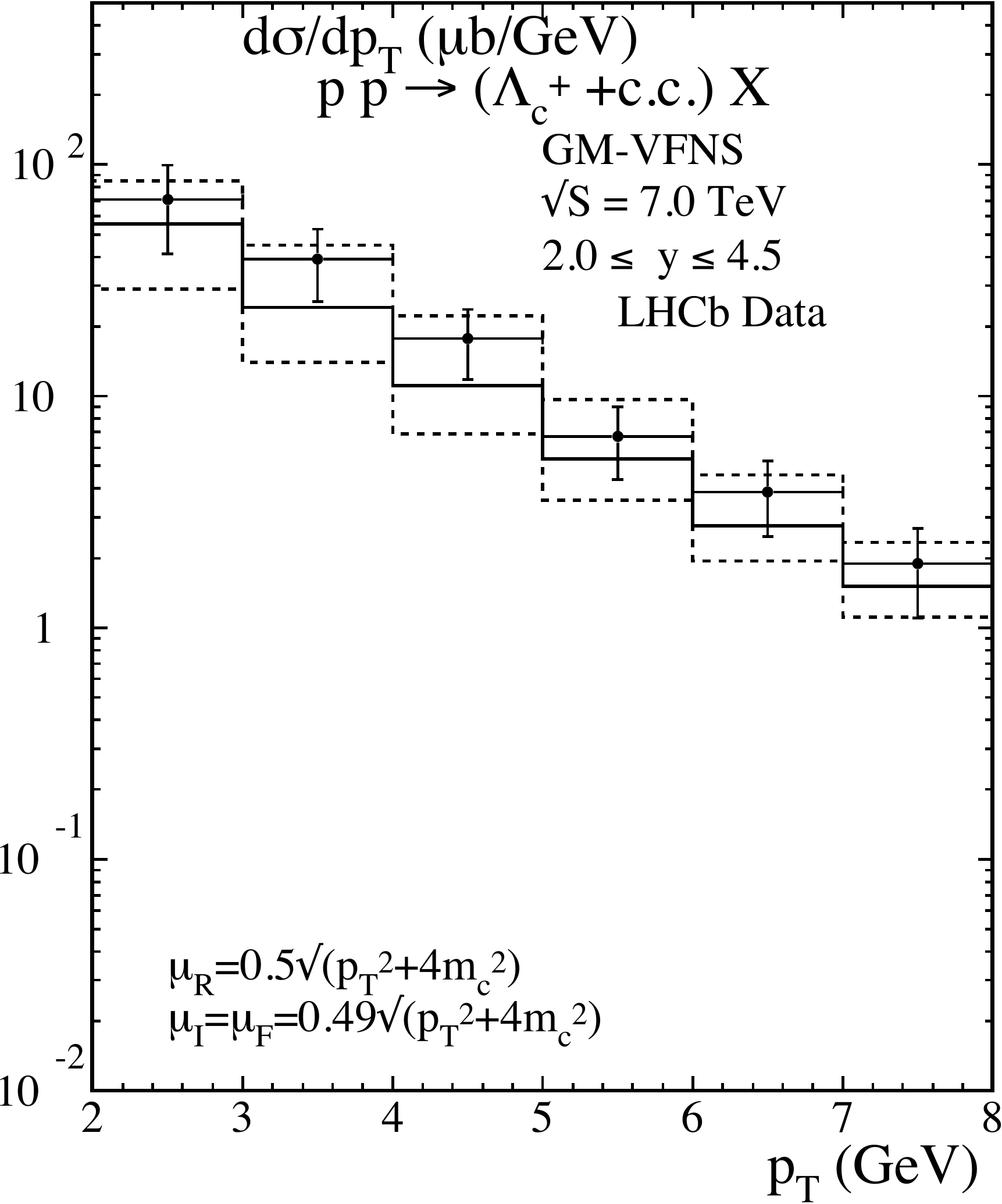}
~~~\includegraphics[height=0.375\textheight]{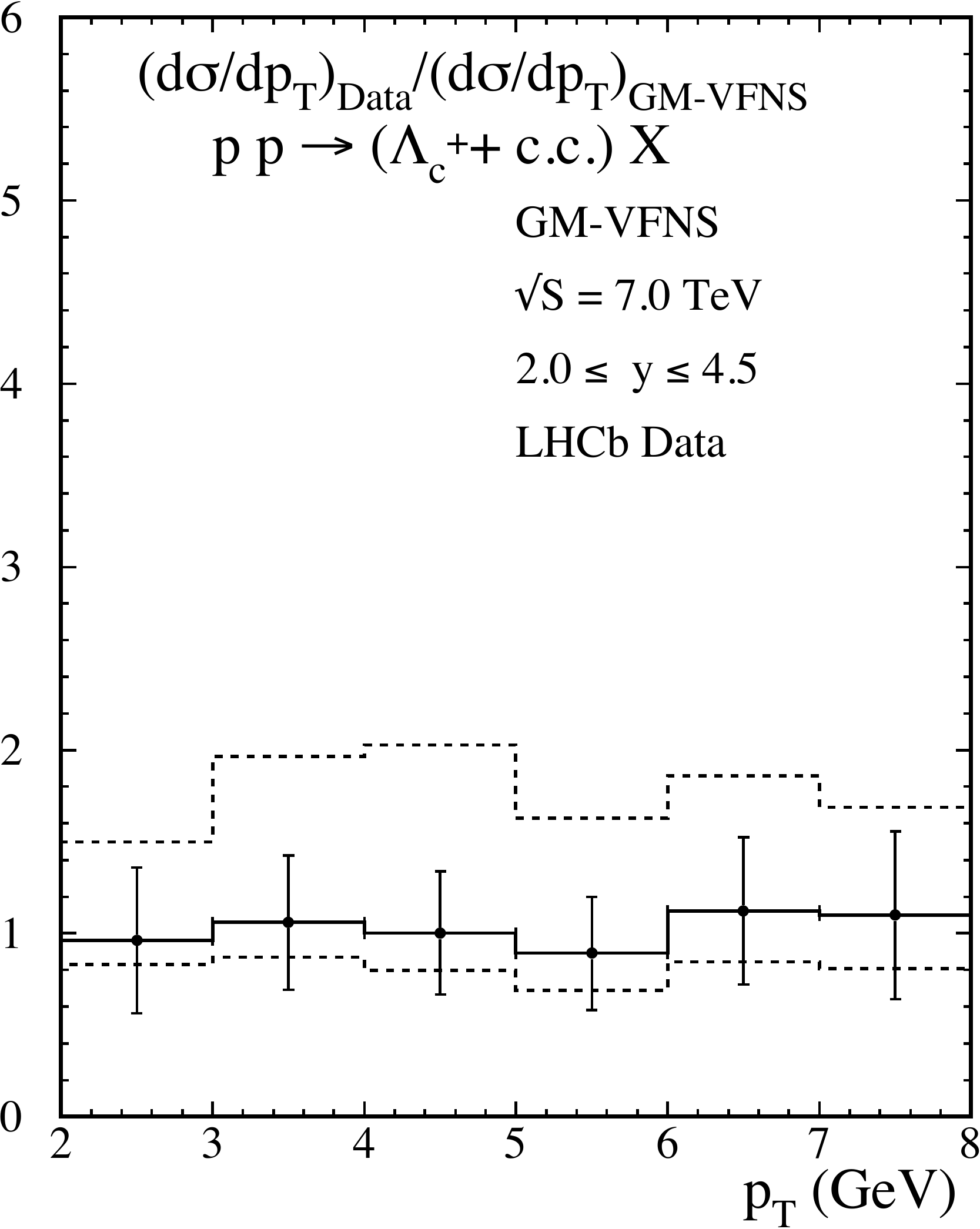}
\end{center}
\caption{
Differential $\Lambda_c^{\pm}$ production cross sections at 
$\sqrt{S} = 7$~TeV as a function of $p_T$ compared with 
LHCb data. The right plot shows the ratio of data over 
theory. The dashed histograms indicate the scale 
uncertainty for $0.5 \leq \xi_R \leq 2.0$.  
\label{fig:1} 
}
\end{figure*}

Our previous calculation of the $\Lambda_c^{\pm}$ production 
cross section was performed with a choice of the renormalization 
and factorization scales which forced us to restrict ourselves 
to large $p_T$. To improve this, we repeat these calculations 
with the same $\Lambda_c^+$ FF of Ref.~\cite{Kniehl:2006mw}, 
but now using the scale parameters as described in 
Sect.~\ref{Sec:input}. The cross sections $d\sigma/dp_T$ 
compared with the LHCb data \cite{Aaij:2013mga} are shown 
in Fig.~\ref{fig:1}, left side. We have rescaled these 
data\footnote{
   This rescaling is needed only for the LHCb data. The data 
   analyses of ALICE and CMS to be discussed below have already 
   used the more recent value of the branching ratio for the 
   $\Lambda_c^{\pm}$ decays.} to be consistent with a more 
recent value for the branching ratio of the $\Lambda_c^+ 
\to \pi^+K^-p$ decay \cite{Tanabashi:2018oca} (factor 0.7874). 
The data thus obtained lie inside the theory uncertainty band 
which is obtained from scale variations for $\mu_R$ using 
scale factors $\xi_R$ ranging from 0.5 to 2.0. The ratio of 
data over theory is presented in Fig.~\ref{fig:1}, right side. 
For the default scale $\xi_R = 1.0$, it agrees with unity inside 
the experimental errors. The error bars for the data are only 
shown for the central curve (full line) corresponding to 
$\xi_R = 1.0$. The histograms with dashed lines correspond 
to $\xi_R = 0.5$ and $\xi_R = 2.0$. 

\begin{figure*}[b!]
\begin{center}
\includegraphics[height=0.365\textheight]{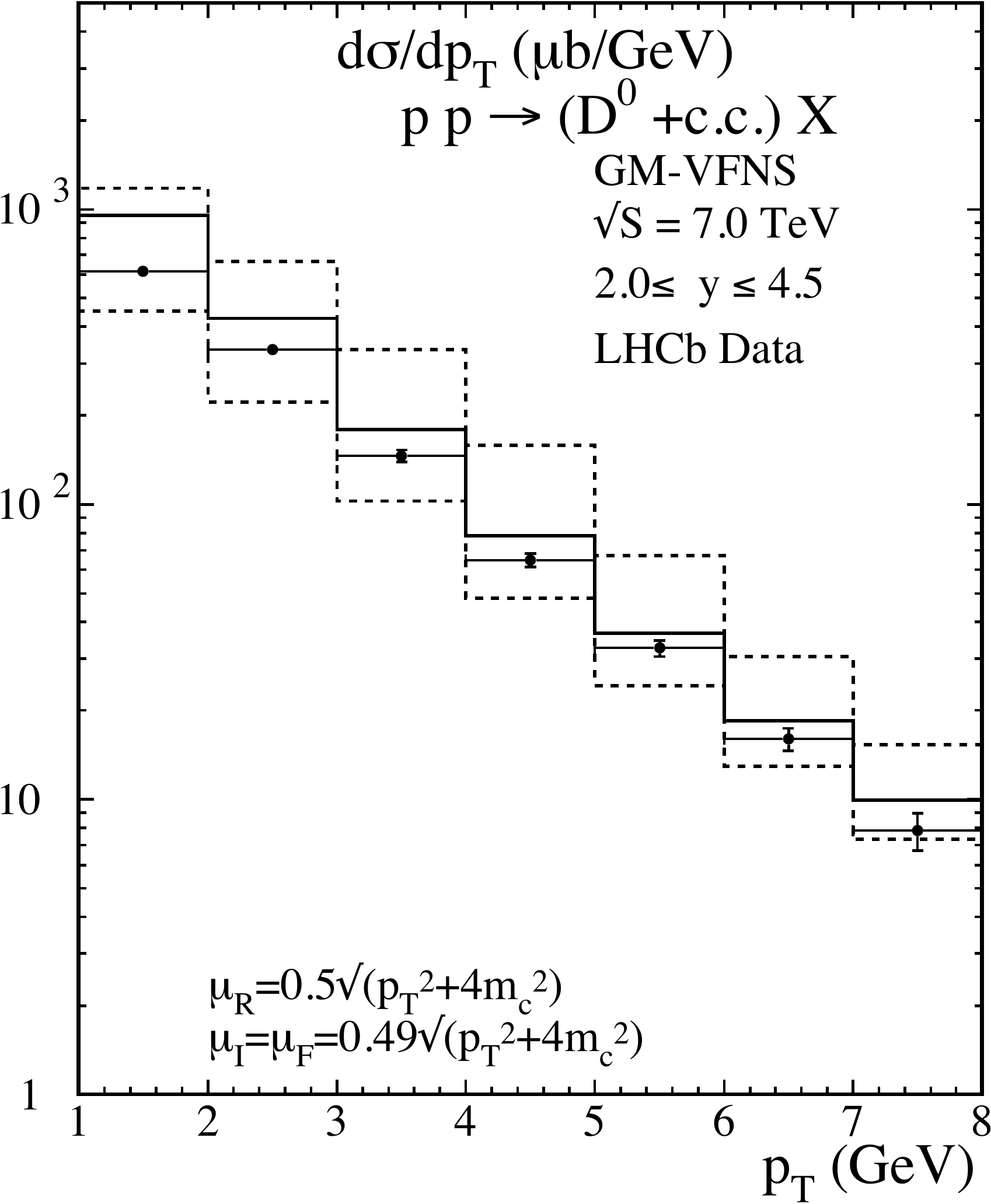}
~~~\includegraphics[height=0.37\textheight]{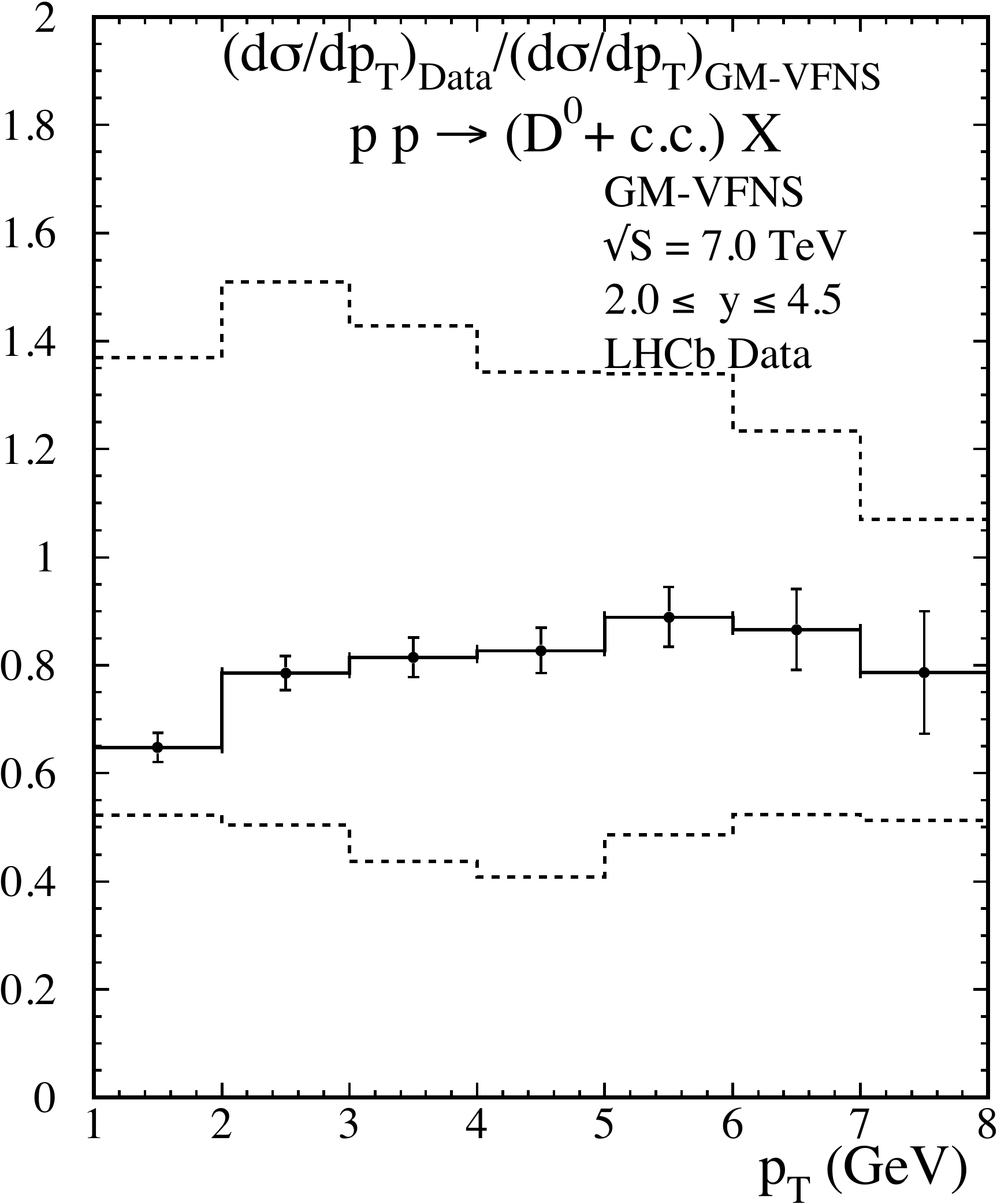}
\end{center}
\caption{ 
Differential $D^0$ production cross sections at 
$\sqrt{S} = 7$~TeV as a function of $p_T$ compared with 
LHCb data. The right plot shows the ratio of data over 
theory. The dashed histograms indicate the scale 
uncertainty for $0.5 \leq \xi_R \leq 2.0$. 
\label{fig:2} 
}
\end{figure*}

\begin{figure*}[b!]
\begin{center}
\includegraphics[height=0.37\textheight]{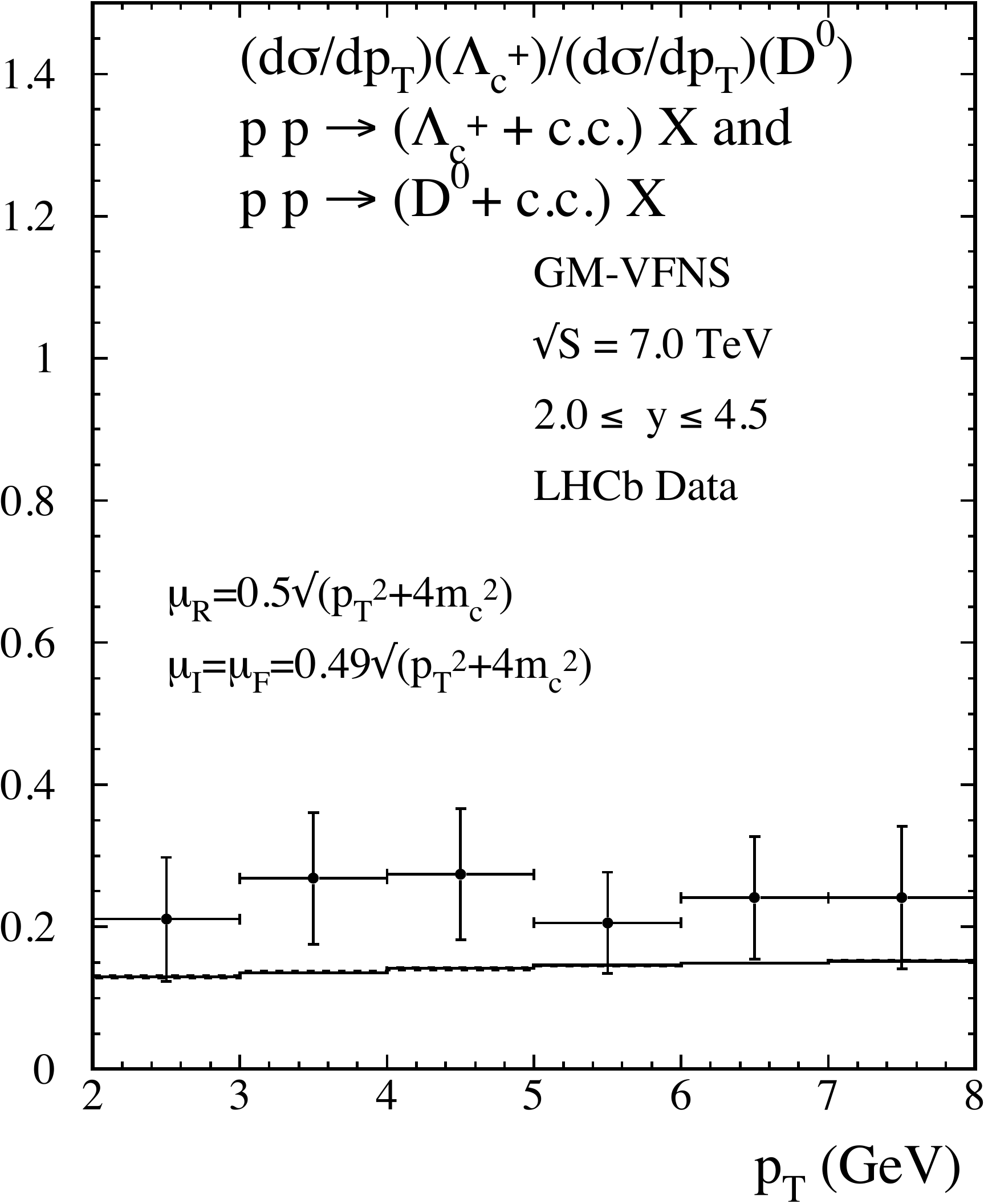}
\end{center}
\caption{ 
$\Lambda_c^{\pm}$ to $D^0$ ratio of production cross sections 
at $\sqrt{S} = 7$~TeV as a function of $p_T$ compared with 
LHCb data. 
\label{fig:3} 
}
\end{figure*}
 
In order to obtain ratios of $\Lambda_c^+$ over $D^0$ 
production, we calculate the cross section for inclusive 
$D^0$ production with the same kinematical conditions and 
with the same choice of scales $\mu_F$ and $\mu_R$ as for 
$\Lambda_c^{\pm}$. They include the inclusive production of 
both charge-conjugate states, $D^0 + c. c.$, as given in 
the LHCb publication \cite{Aaij:2013mga}. The predictions are 
compared with the data \cite{Aaij:2013mga} for $\sqrt{S} = 
7$~TeV. We find agreement within the theory uncertainty 
band given by the scale variation (see Fig.~\ref{fig:2}, left 
side). The ratio of data over theory, shown in the right panel 
of Fig.~\ref{fig:2}, is approximately $0.8$. Experimental 
uncertainties are rather small, and the deviation of this 
ratio from unity is quite significant, but agrees with theory 
within the larger theory uncertainties. Using these results, 
we can now calculate the ratio of $\Lambda_c^+$ and $D^0$ 
cross sections as a function of $p_T$. The result is shown 
in Fig.~\ref{fig:3}. The predicted ratio is approximately 
equal to 0.15 and below the experimental value of $\simeq 
0.2$ by about one standard deviation of the experimental 
errors. One should note that the scale dependence of the 
theory prediction cancels to a good degree in the ratio of 
cross sections. The dependence on PDF uncertainties is 
expected to be much smaller than the scale dependence 
\cite{Kramer:2018vde} and would also cancel to some extent 
in the ratio of cross sections. 

\begin{figure*}[b!]
\begin{center}
\includegraphics[height=0.37\textheight]{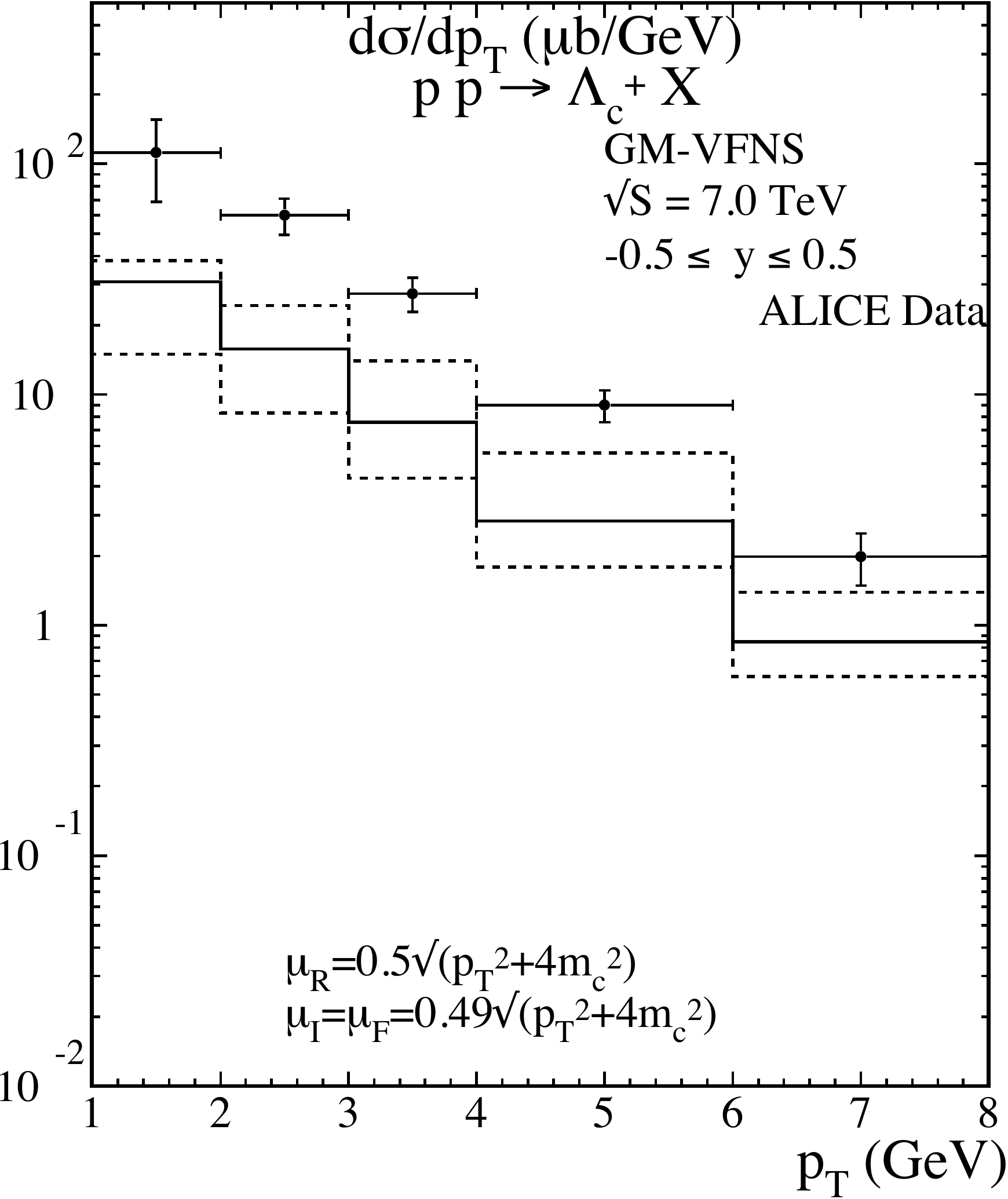}
\raisebox{1mm}{
~~~\includegraphics[height=0.37\textheight]{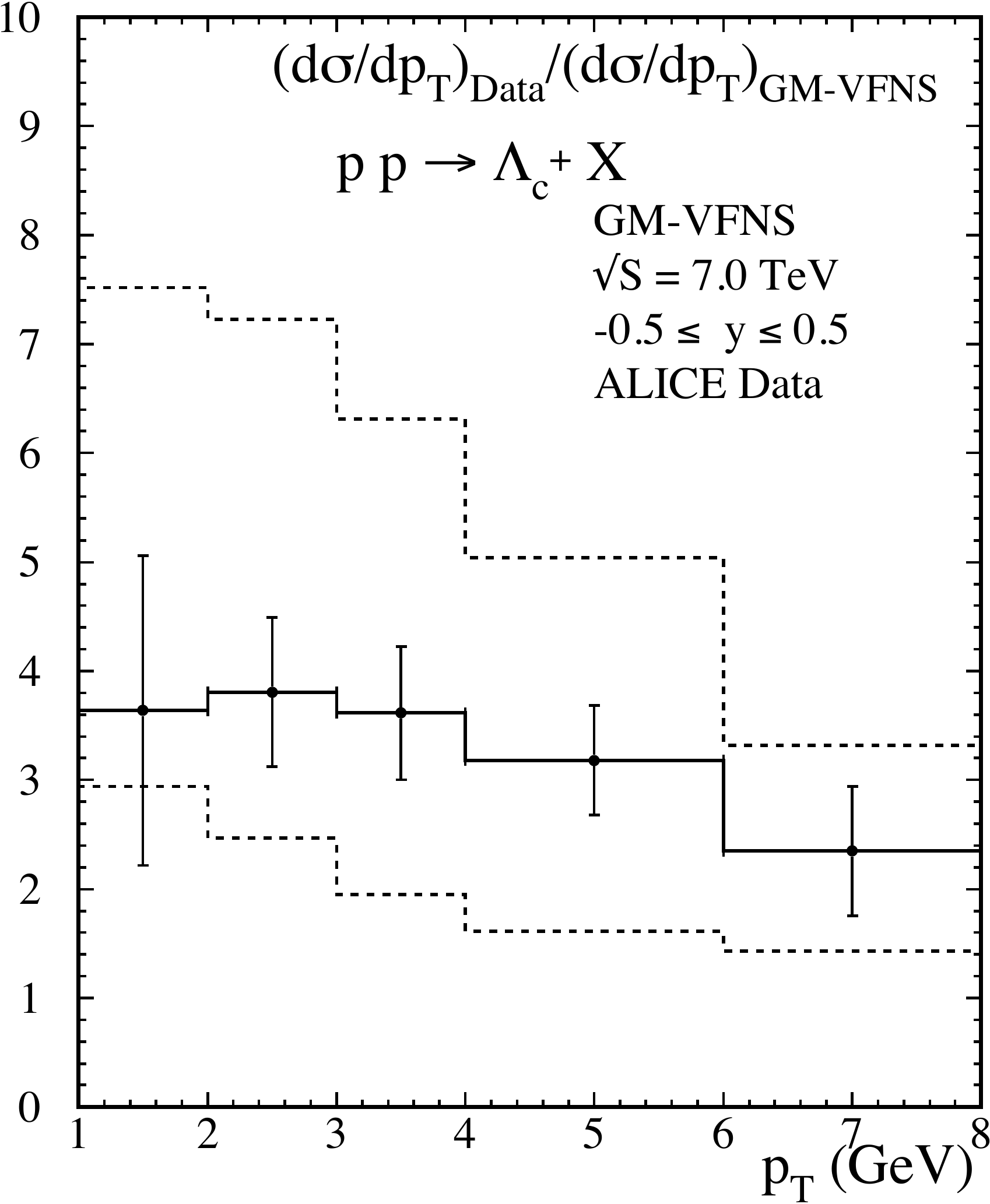}
}
\end{center}
\caption{
Differential $\Lambda_c^{\pm}$ production cross sections at 
$\sqrt{S} = 7$~TeV as a function of $p_T$ compared with 
ALICE data. The right plot shows the ratio of data over 
theory. The dashed histograms indicate the 
scale uncertainty for $0.5 \leq \xi_R \leq 2.0$. 
\label{fig:4} 
}
\end{figure*}

We repeat these calculations to compare with ALICE data 
\cite{Acharya:2017kfy}. These data have been obtained 
for central production $|y| \leq 0.5$ at $\sqrt{S}=7$~TeV 
and in five $p_T$ bins between $1$~GeV and $8$~GeV. One 
should note that these data are for inclusive $\Lambda_c^{\pm}$ 
production without including charge-conjugate states, in 
contrast to data from the LHCb collaboration. We choose the 
prescription of Eq.~(\ref{eq:scales}) to fix the renormalization 
and factorization scales. The results are shown in 
Fig.~\ref{fig:4}, left side, and compared with the ALICE 
measurements \cite{Acharya:2017kfy}. For all five $p_T$ bins 
the data are larger than our predictions and outside the 
theory error band due to scale variations. The ratio of data 
over GM-VFNS results, shown in Fig.~\ref{fig:4}, right side, 
ranges between $3.6 \pm 1.4$ (in the bin with smallest $p_T$) 
and $2.4 \pm 0.6$ (in the bin with largest $p_T$). This 
agrees with the comparison shown in Ref.~\cite{Acharya:2017kfy}, 
where predictions with a different choice of scales have been 
used. These results show, as already stated in the ALICE 
publication \cite{Acharya:2017kfy}, that the measured 
cross sections in the central $|y|$ region are much larger 
than the predictions. In contrast, as shown above, the 
LHCb results in the forward $y$ region are compatible 
with predictions based on the $\Lambda_c^+$ FF from 
Ref.~\cite{Kniehl:2006mw}. 

\begin{figure*}[b!]
\begin{center}
\includegraphics[height=0.37\textheight]{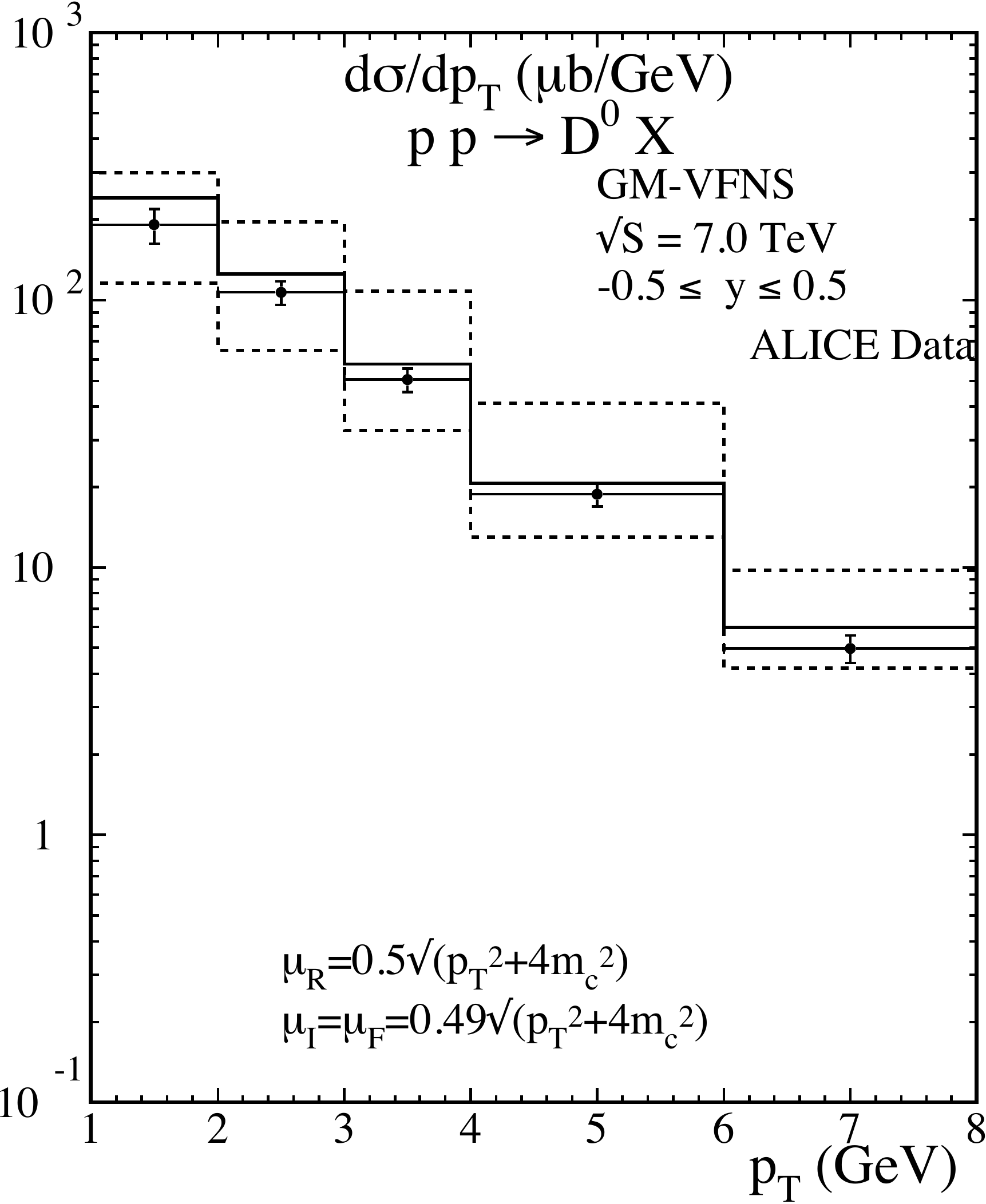}
\raisebox{-1.3mm}{
~~~\includegraphics[height=0.37\textheight]{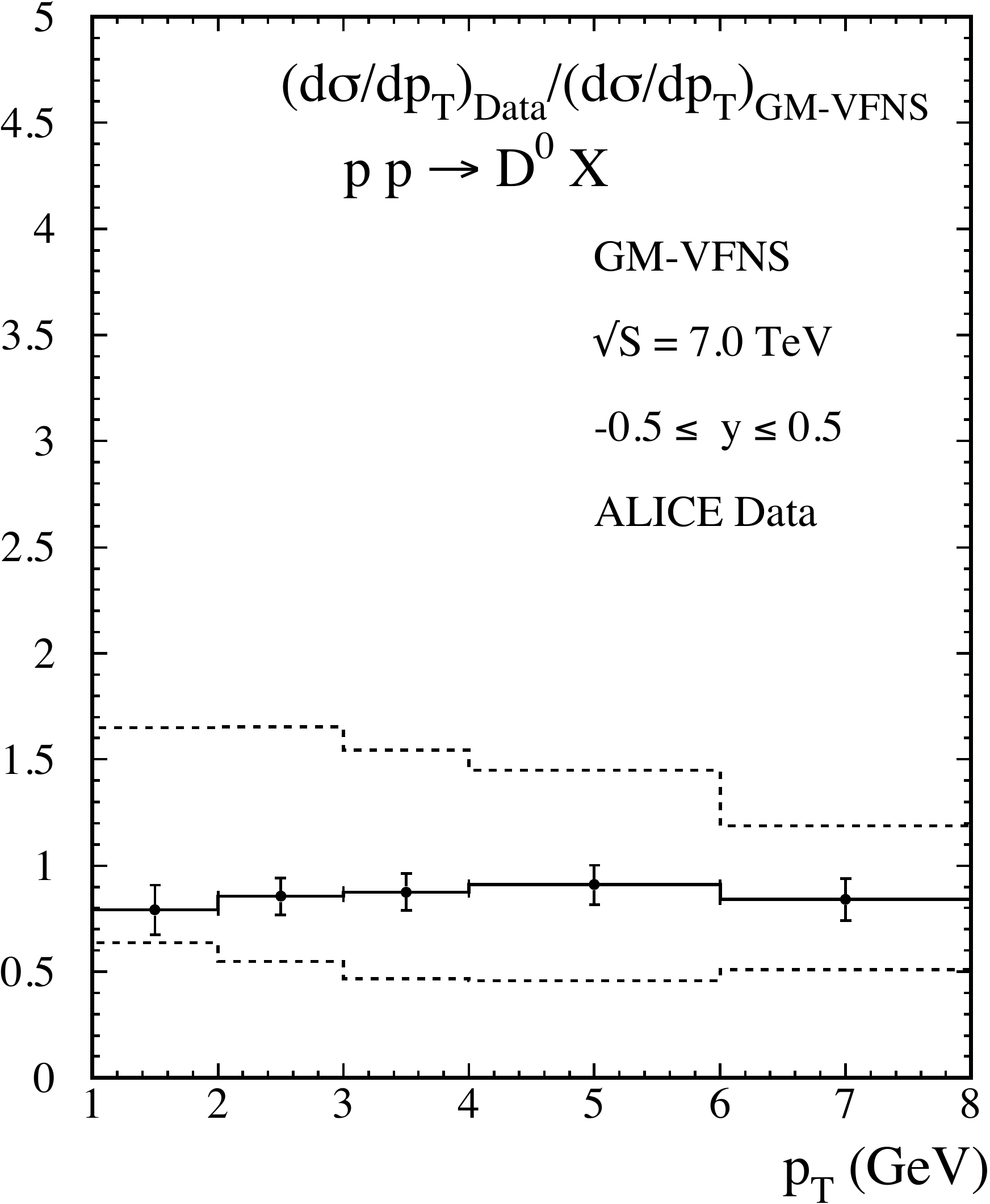}
}
\end{center}
\caption{
Differential $D^0$ production cross sections at 
$\sqrt{S} = 7$~TeV as a function of $p_T$ compared with 
ALICE data. The right plot shows the ratio of data over 
theory. The dashed histograms indicate the scale 
uncertainty for $0.5 \leq \xi_R \leq 2.0$. 
\label{fig:5} 
}
\end{figure*}

\begin{figure*}[t!]
\begin{center}
\includegraphics[height=0.37\textheight]{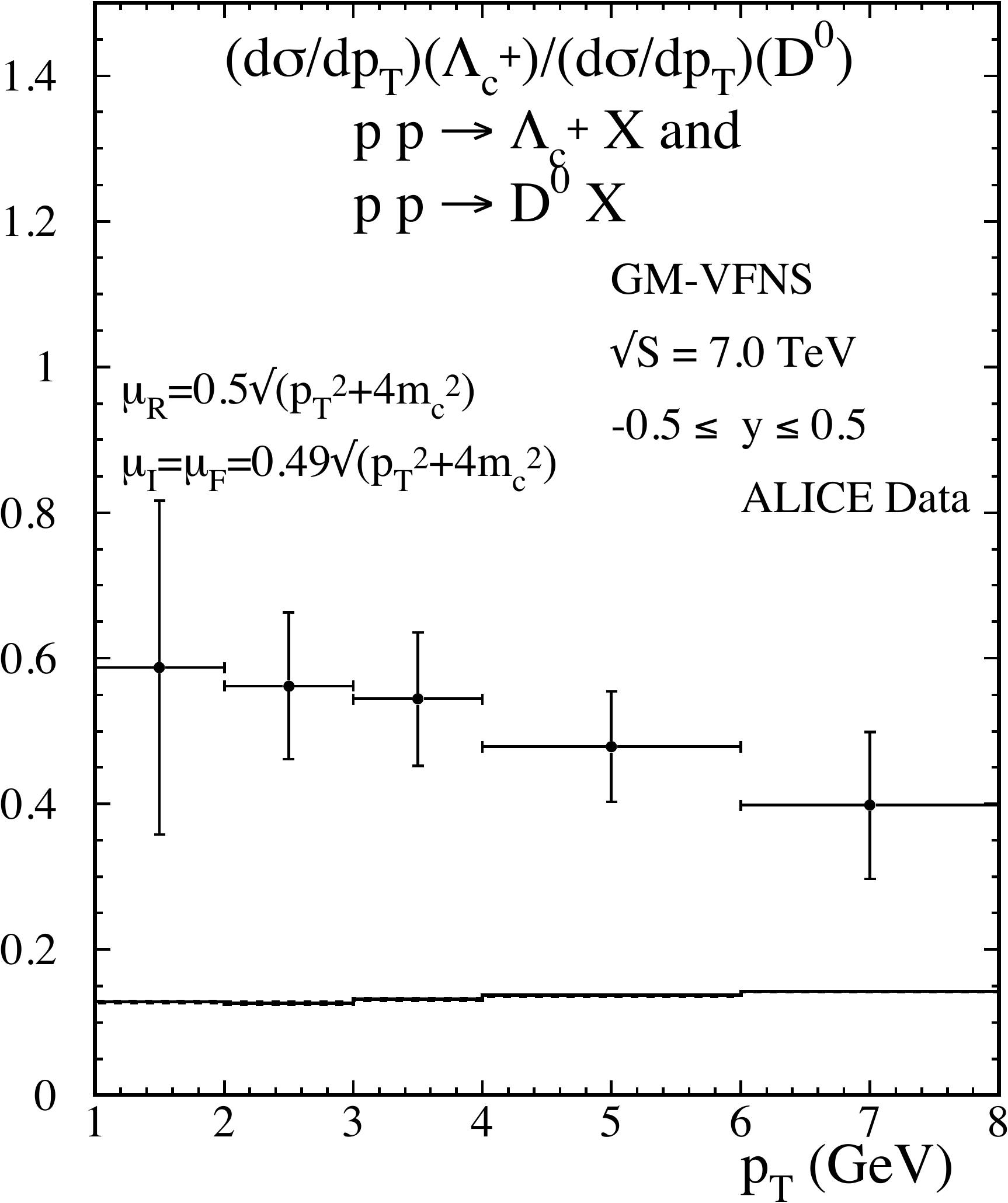}
\end{center}
\caption{
$\Lambda_c^+$ to $D^0$ ratios of production cross sections 
at $\sqrt{S} = 7$~TeV as a function of $p_T$ compared with 
ALICE data. 
\label{fig:6} 
}
\end{figure*}

To obtain the $\Lambda_c^+ / D^0$ ratio, we need the inclusive 
$D^0$ cross sections, again for the same bin sizes and with 
the same choice of scales. Results are shown in Fig.~\ref{fig:5}, 
left side, and compared with ALICE data taken from 
Refs.~\cite{Acharya:2017jgo,Adam:2016ich}. The experimental 
data fall inside the uncertainty band due to $\xi_R$ variation 
in the range $1/2 < \xi_R < 2$, where the lower value of $\xi_R$ 
leads to the maximum prediction. The full-line histogram is for 
the default choice $\xi_R = 1.0$. It agrees fairly well with 
the experimental data. The ratio of data over theory is 
shown in Fig.~\ref{fig:5}, right side. This ratio is 
approximately equal to unity. The comparison of data and theory 
for the $\Lambda_c^+/D^0$ cross section ratio is shown in 
Fig.~\ref{fig:6}. One can see clearly a disagreement between 
data and prediction. Experimental values for this ratio are 
found to range between 0.6 (at small $p_T$) and 0.4 (at large 
$p_T$). They exceed the theoretical prediction by factors of 
approximately 4.0 at small $p_T$ and 2.7 at large $p_T$. 
Theory predicts a value of $\simeq 0.15$, which is independent 
of the scale choice. Comparing Figs.~\ref{fig:4} and \ref{fig:5}, 
it is obvious that the discrepancy seen in the $\Lambda_c^+/D^0$ 
ratio originates solely from the $\Lambda_c^{\pm}$ cross section. 

\begin{figure*}[t!]
\begin{center}
\includegraphics[height=0.37\textheight]{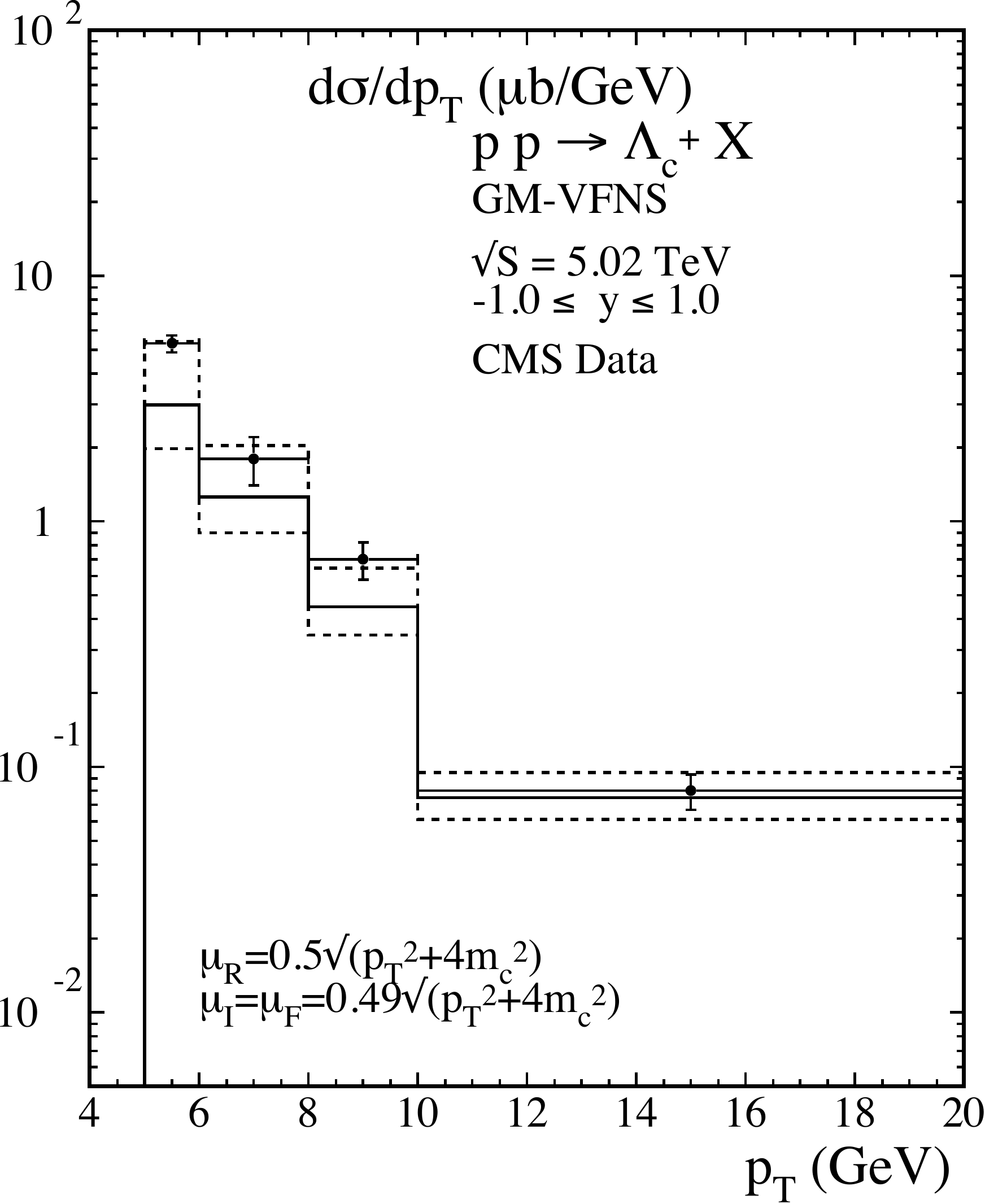}
\raisebox{-0.8mm}{
\includegraphics[height=0.37\textheight]{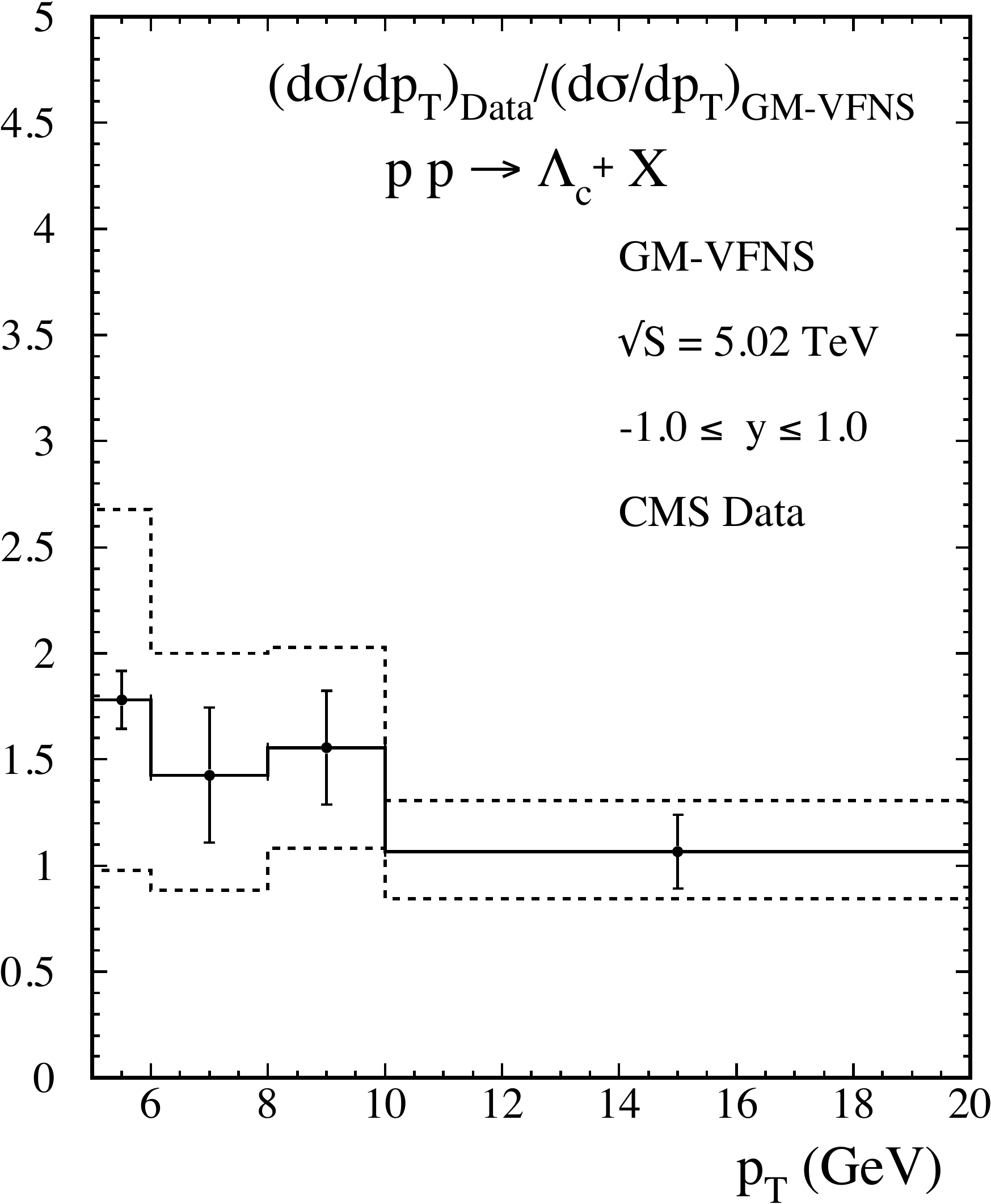}
}
\end{center}
\caption{
Differential $\Lambda_c^{\pm}$ production cross sections at 
$\sqrt{S} = 5.02$~TeV as a function of $p_T$ compared with 
CMS data. The right plot shows the ratio of data over theory. 
The dashed histograms indicate the scale uncertainty for 
$0.5 \leq \xi_R \leq 2.0$. 
\label{fig:7} 
}
\end{figure*}

\begin{figure*}[h!]
\begin{center}
\includegraphics[height=0.36\textheight]{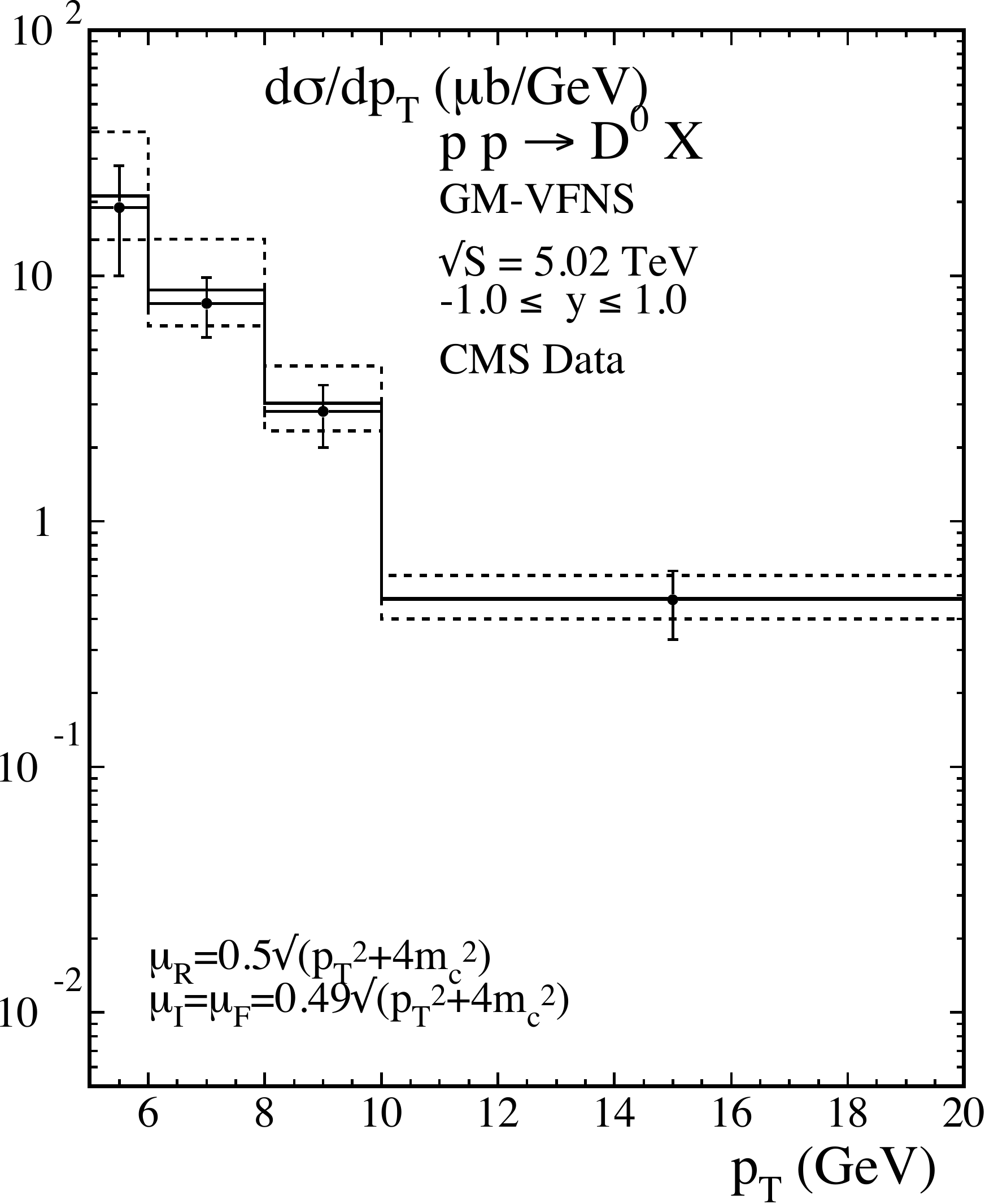}
\raisebox{-0.8mm}{
\includegraphics[height=0.36\textheight]{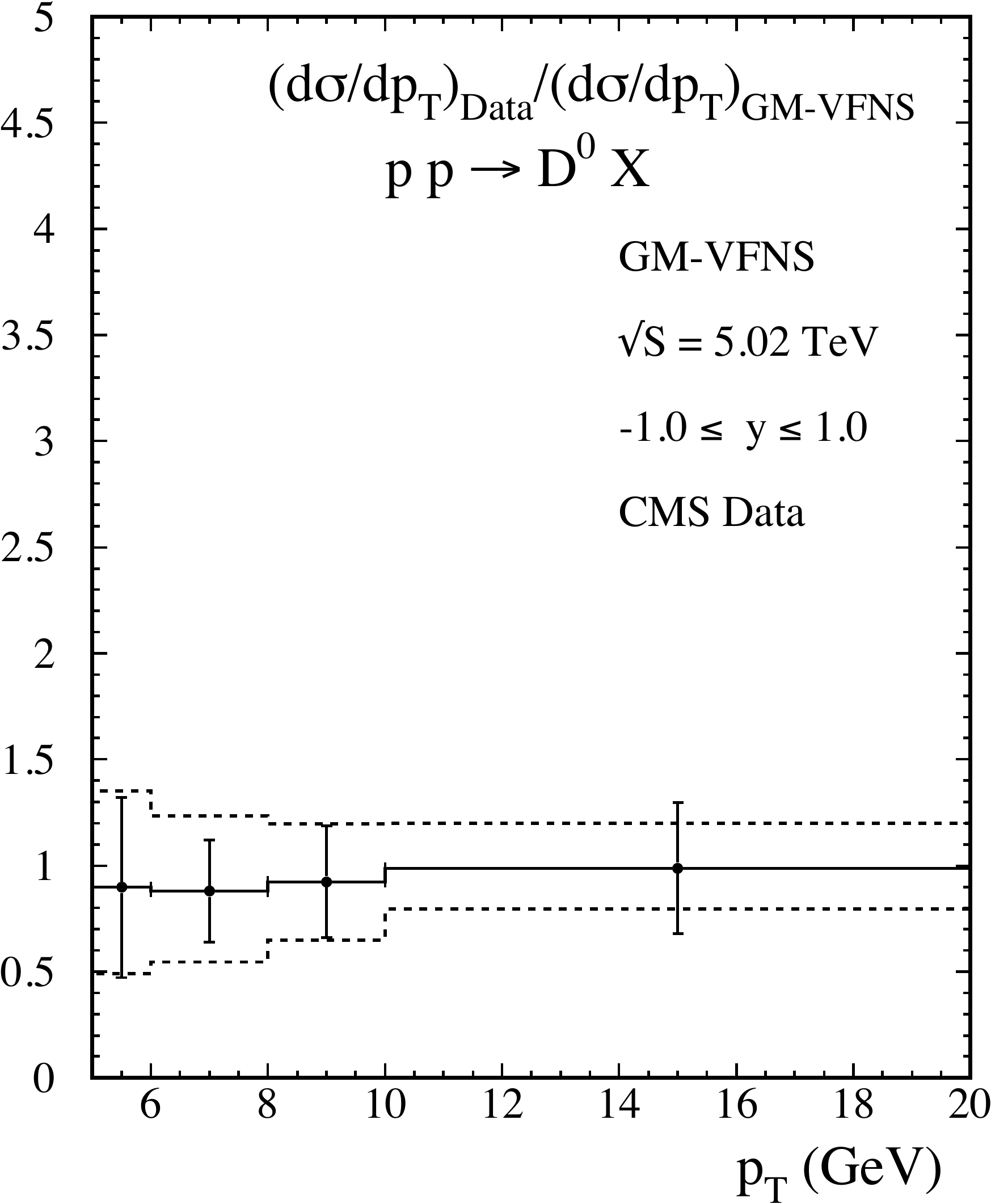}
}
\end{center}
\caption{
Differential $D^0$ production cross sections at 
$\sqrt{S} = 5.02$~TeV as a function of $p_T$ compared with 
CMS data. The right plot shows the ratio of data over theory. 
The dashed histograms indicate the scale uncertainty for 
$0.5 \leq \xi_R \leq 2.0$. 
\label{fig:8} 
}
\end{figure*}

\begin{figure*}[b!]
\begin{center}
\includegraphics[height=0.36\textheight]{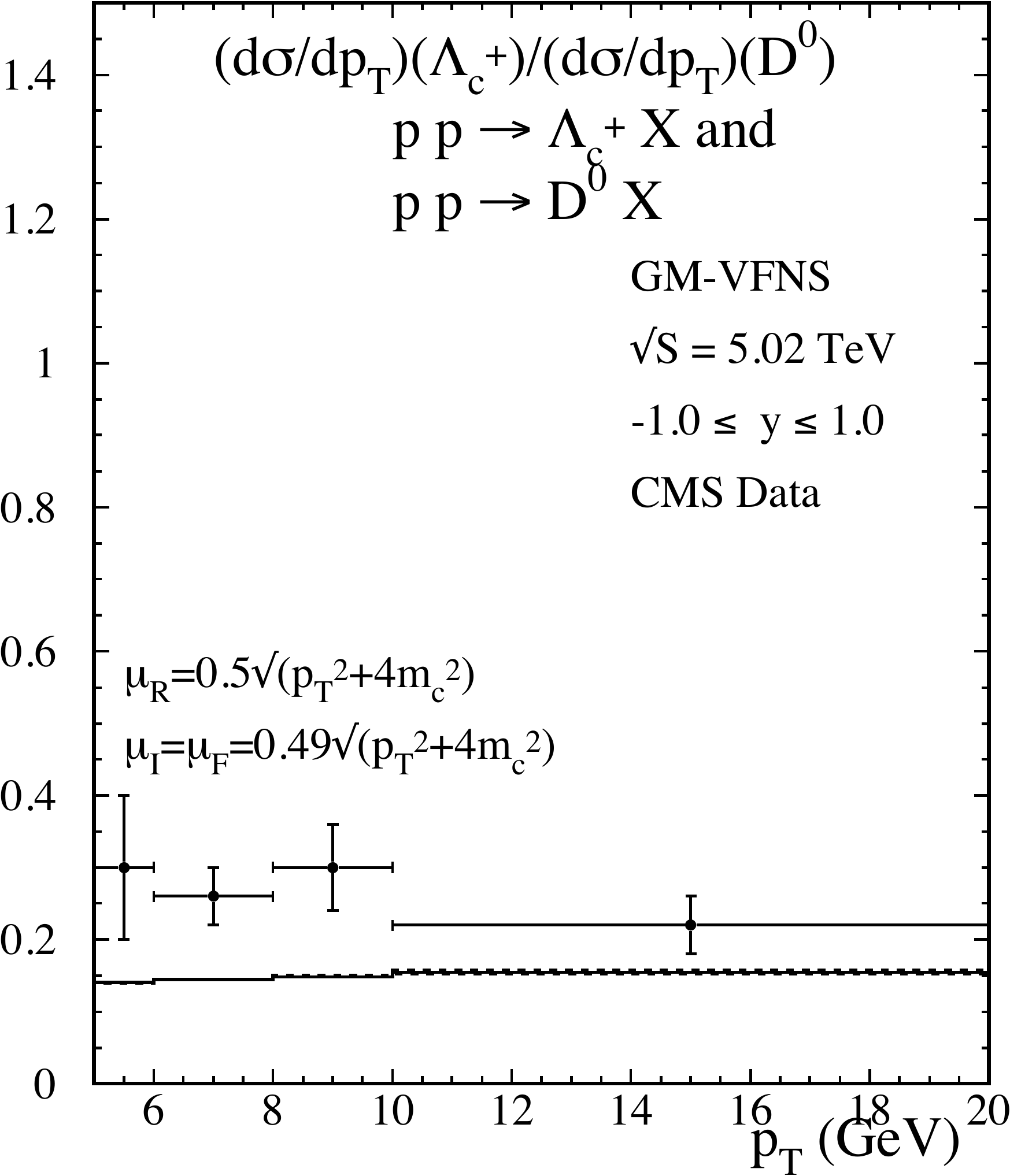}
\end{center}
\caption{
$\Lambda_c^+$ to $D^0$ ratio of production cross sections 
at $\sqrt{S} = 5.02$~TeV as a function of $p_T$ compared 
with CMS data. 
\label{fig:9} 
}
\end{figure*}

Finally, we present results for $\Lambda_c^{\pm}$ and $D^0$ 
production at $\sqrt{S}=5.02$~TeV, which we compare with 
the recent CMS measurements \cite{Sirunyan:2019fnc}. Data 
from CMS are available for $d\sigma/dp_T$ in four $p_T$ 
bins in the range between 5 and 20~GeV, and in the rapidity 
interval $|y| < 1.0$. This kinematic range is similar to 
the one of the ALICE measurements \cite{Acharya:2017kfy}. 
Both cover the central rapidity range, somewhat larger 
in the case of CMS ($|y|<1.0$) than in the case of ALICE 
($|y|<0.5$). Our results are shown in Fig.~\ref{fig:7}, 
left side, and compared with the four data points from CMS 
\cite{Sirunyan:2019fnc}.  The ratio of data over GM-VFNS 
predictions is presented in Fig.~\ref{fig:7}, right side, 
and agrees with unity at the lower border of the uncertainty 
band due to scale variations, i.e.\ within theory errors. 
The results of $d\sigma/dp_T$ for $D^0$ production in the 
same $p_T$ bins are shown in Fig.~\ref{fig:8}, left side. 
Our results are compared with CMS data, which we have taken 
from the corresponding figure in Ref.~\cite{Sirunyan:2017xss}. 
We find a very good agreement between data and the calculation 
using default scales. The ratio of data over theory shown in 
Fig.~\ref{fig:8}, right side is equal to unity, as expected. 
The ratio of $d\sigma/dp_T$ for $\Lambda_c^+$ over $D^0$ 
production is shown in Fig.~\ref{fig:9}. Theory predicts a 
ratio of $\simeq 0.15$, a result similar to the one obtained 
for the LHCb kinematic range shown in Fig.~\ref{fig:5}, 
i.e.\ theory does not predict for this ratio a strong 
dependence on the rapidity range. The CMS data for the 
$\Lambda_c^+/D^0$ ratio shown in Fig.~\ref{fig:9} is 
approximately $0.3$, only a factor two larger than the 
theoretical result. The data point for the bin 
$6<p_T<8$~GeV in Fig.~\ref{fig:9} can be compared with 
a data point in the same $p_T$ bin from ALICE, see 
Fig.~\ref{fig:6}. The two data differ only by the 
different sizes of the $y$ coverage. The ALICE point 
is found at a value of $0.4 \pm 0.1$, whereas the data 
point in Fig.~\ref{fig:9} from CMS is at $0.265 \pm 0.112$, 
somewhat below the ALICE value. It is unclear whether this 
difference can be attributed to the smaller $|y|$ range 
for the ALICE point. 

Comparing the theory predictions for the ratio of cross 
sections for $\Lambda_c^{\pm}$ and $D^0$ production shown 
in Figs.~\ref{fig:3}, \ref{fig:6} and \ref{fig:9}, we can 
see that they are very similar; there is neither an indication 
for a strong dependence on rapidity, nor on the range 
of $p_T$ values which have been investigated by the three 
experiments, LHCb ($2 < p_T < 8$~GeV), ALICE ($1 < p_T < 
8$~GeV) and CMS ($5 < p_T < 20$~GeV). We should expect 
that also the measured $\Lambda_c^+/D^0$ ratios in the 
three experiments should be equal. This is, however, not 
the case. The data for the $\Lambda_c^+/D^0$ ratio  
measured by LHCb and CMS are almost equal inside 
their large experimental uncertainties, $\simeq 0.3$ 
for both experiments. In contrast, the values found by 
ALICE are larger. This discrepancy is most significant 
at small $p_T$, where the ALICE value of the ratio is 
$\sim 0.6$, while their value $\sim 0.4$ at large $p_T$ 
comes closer to the results from the LHCb and CMS 
collaborations. 

The situation is different in the case of $b$-hadron 
production. In a previous work \cite{Kramer:2018rgb}
we have compared predictions for $\Lambda_b^0$ production 
with data from the LHCb collaboration. At small $p_T$, the 
ratio of $\Lambda_b^0$ to $B^0$ production cross sections 
was in agreement with theory, which is essentially determined 
by the ratio of the $b \to \Lambda_b^0$ to $b \to B$ 
fragmentation fractions, $f_{\Lambda_b} / f_d \simeq 0.25$. 
Only at $p_T \gsim 10$~GeV, the LHCb data of this ratio 
decrease and reach the value 0.15 at $p_T \simeq 20$~GeV, 
which is about 2 to $3\sigma$ below the expected value. 

\section{New fit of the $\Lambda_c^+$ FF}

The old FF parametrization used in the previous section 
was based on a fit to OPAL data \cite{Alexander:1996wy} 
including only 4 points with rather large uncertainties. 
Here we describe a new fit including Belle data at $\sqrt{S} 
= 10.52$~GeV \cite{Niiyama:2017wpp}. The Belle data set is 
much more precise and contains more points. Only 35 of the 
available total number of 42 points will be used in our fit, 
since we have to exclude data at small values of the scaling 
variable $x$, where theory is not reliable without taking 
resummation of soft-gluon logarithms into account. 

The combination of old OPAL data with the more recent data 
from Belle requires special care, since the two experiments 
have based their analyses on the observation of the decay 
$\Lambda_c^+ \to \pi^+K^-p$ for which different branching 
ratios have been used. OPAL has used the 1996 value 
Br$(\Lambda_c^+ \to \pi^+K^-p) = 0.044$
\cite{Alexander:1996wy}, whereas the analysis of the 
Belle measurements relies on the world average 
Br$(\Lambda_c^+ \to \pi^+K^-p) =  0.0635$ from the 2016 
Review of Particle Physics \cite{Patrignani:2016xqp}. 
We, therefore, re-scale the OPAL cross sections by the factor 
$0.044/0.0635 =  0.6929$. The branching fractions for the 
transitions $c \to \Lambda_c^+$ and $b \to \Lambda_c^+$ 
are correspondingly reduced by this factor. 

The strategy for constructing the $\Lambda_c^+$ FF is 
the same as in our previous work with T.~Kneesch 
for $D$-meson FFs \cite{Kneesch:2007ey}. The 
$e^+e^-$ annihilation cross sections are calculated at 
NLO with corrections for non-zero charm and bottom masses. 
Corrections for the finite mass of the charmed baryon 
$\Lambda_c^{\pm}$ are also taken into account. We parametrize 
the $x$-dependence of the $c$- and $b$-quark FFs at their 
respective starting scales as suggested by Bowler 
\cite{Bowler:1981sb}: 
\begin{equation} 
D_Q(x,\mu_0) = N x^{-(1+\gamma)^2} (1-x)^a e^{-\gamma^2/x} 
\, , 
\label{eq:bowler}
\end{equation} 
with three parameters $N$, $a$ and $\gamma$ for each of the 
quarks $Q =c$, $b$. The starting scales are $\mu_0 = 1.5$~GeV 
for $Q=c$ and $\mu_0 = 5$~GeV for $Q=b$. The fitting 
procedure is as follows. At the starting scale $\mu_0 = 
1.5$~GeV the $c$-quark FF is taken to be of the form given 
in Eq.~(\ref{eq:bowler}), while the FFs of the light quarks 
$u$, $d$, $s$ and the gluon are set equal to zero. Then these 
FFs are evolved to higher scales using the DGLAP evolution 
equations (see, for example, Eq.~(10) in 
Ref.~\cite{Kneesch:2007ey}) at NLO with $n_f=4$ active quark 
flavors and a given value of $\Lambda_{\overline{\rm MS}}^{(4)}$. 
When the scale reaches the bottom threshold at $\mu_F = m_b 
= 5$~GeV, the bottom flavour is activated and its FF is 
introduced in the Bowler form of Eq.~(\ref{eq:bowler}). 
The evolution to higher scales is performed with $n_f = 5$ 
and the value $\Lambda_{\overline{\rm MS}}^{(5)}$ is properly 
adjusted to match the value of $\Lambda_{\overline{\rm MS}}^{(4)}$.
We also note that, in Ref.~\cite{Niiyama:2017wpp}, the Belle 
collaboration has provided data which are corrected for 
radiative effects. We use these data and thus do not have 
to apply radiative corrections ourselves, as described in 
Ref.~\cite{Kneesch:2007ey}. 

Similarly to the work in Ref.~\cite{Kneesch:2007ey}, we 
perform three different fits. First, FFs are determined 
using separately the $B$-factory data (Belle fit) and the 
rescaled $Z$-factory data (OPAL fit). Then we obtain a common 
fit (global fit) combining the two data sets. 

\begin{figure*}[t!]
\begin{center}
\includegraphics[width=0.48\linewidth]{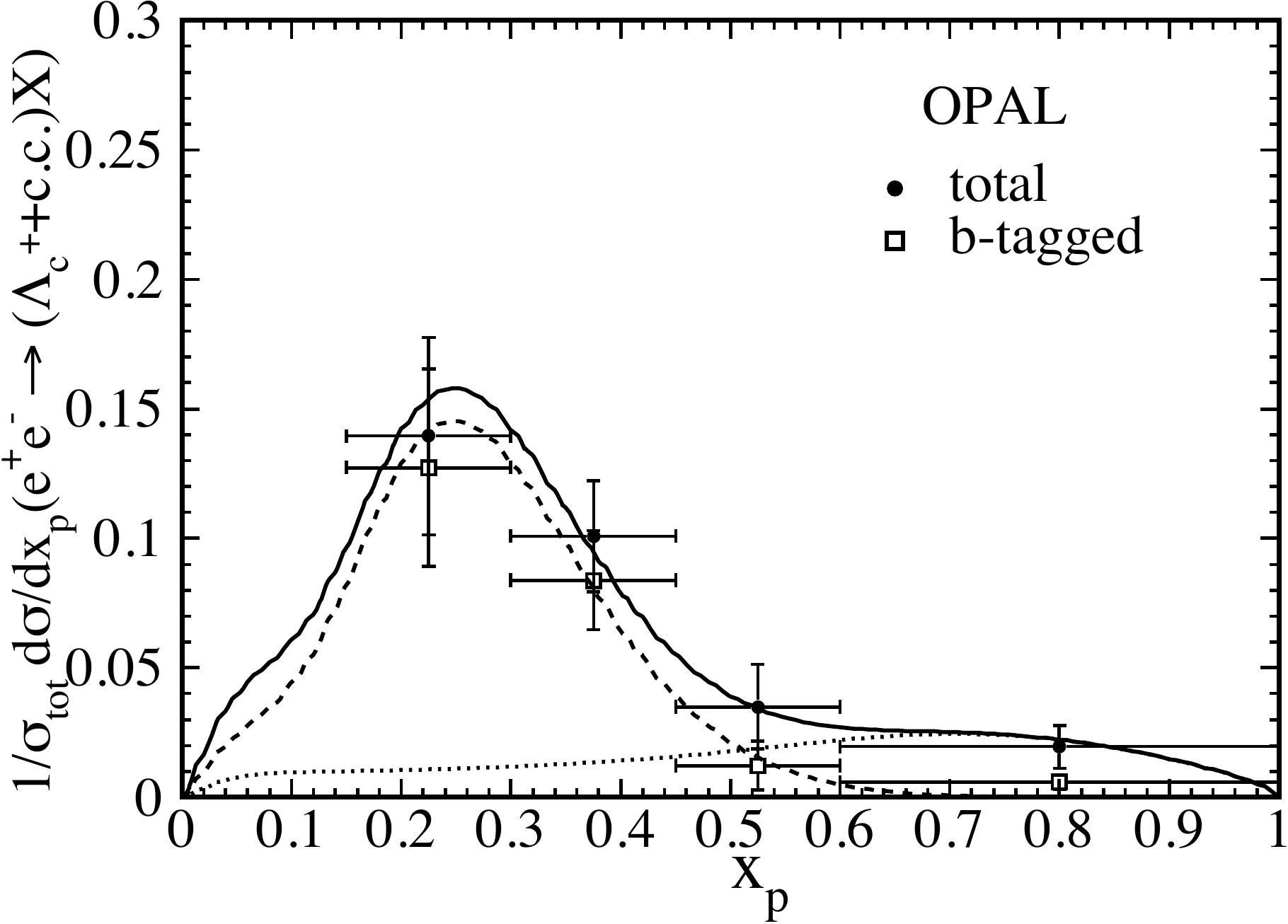}
~~\includegraphics[width=0.48\linewidth]{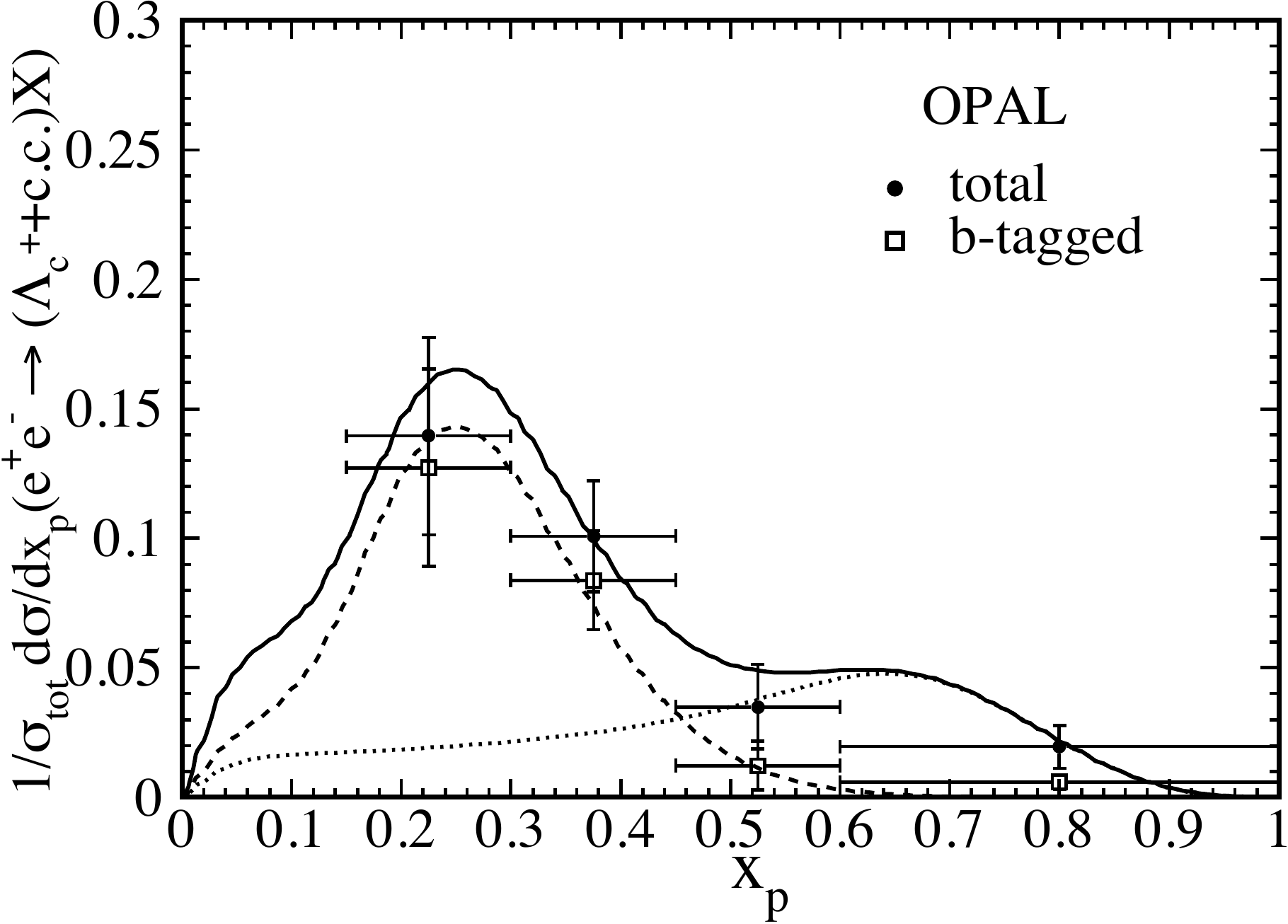}
\end{center}
\caption{
OPAL fit (left) and global fit (right) compared 
with OPAL data. The dashed line shows the contribution 
originating from $b$-quark production, the dotted line 
describes the $c$ component and the full line is the 
sum of both contributions to the normalized production 
cross section. 
\label{fig:13} 
}
\end{figure*}

We start by updating the FF fit to the OPAL data using 
the rescaled cross sections to match the up-to-date value 
of Br$(\Lambda_c^+ \to \pi^+K^-p)$. There are two data 
sets coming from OPAL \cite{Alexander:1996wy}: one sample 
includes only $\Lambda_c^{\pm}$ baryons produced in the decays 
of $b$ hadrons from $Z \to b\bar{b}$ (denoted $b$-tagged); 
a second sample includes in addition $\Lambda_c^{\pm}$ baryons 
from the direct production in $Z \to c \bar{c}$ events and 
from light-quark and gluon fragmentation (denoted total). 
We determine the FFs for $c \to \Lambda_c^+$ and $b \to 
\Lambda_c^+$ in a common fit. The resulting values of the 
fit parameters and the $\chi^2$ per degree of freedom 
are given in Table \ref{tab:fitpars} in the column denoted 
``OPAL''. The quality of the fit may be judged from 
Fig.~\ref{fig:13} (left). The fragmentation fractions for 
$c \to \Lambda_c^+$ and $b \to \Lambda_c^+$ resulting from 
this fit at $\mu_F = 10.52$~GeV are listed in 
Table~\ref{tab:fragratios} and the average energy 
fractions in Table \ref{tab:eratios}. Compared with our 
previous fit \cite{Kniehl:2006mw}, these values have changed 
as expected. 

In the Belle data \cite{Niiyama:2017wpp}, contributions 
from $B$-meson decays are excluded, so that the $b \to 
\Lambda_c^+$ FF is not needed in the calculation of 
cross sections. However, the FFs from $c$ and $b$ quarks 
are coupled through the DGLAP evolution. We fix the 
$b \to \Lambda_c^+$ FF using the values of $N_b$, $a_b$ 
and $\gamma_b$ obtained from the OPAL fit. The fit to 
the Belle data yields new values for $N_c$, $a_c$ and 
$\gamma_c$, which are shown in the column denoted ``Belle'' 
in Table \ref{tab:fitpars}. The corresponding values for 
the fragmentation and average energy fractions are given 
in Tables \ref{tab:fragratios} and \ref{tab:eratios}, 
respectively. 

\begin{table}[t!]
  \begin{center}
    \begin{tabular}{llll}
      \hline
        & OPAL & Belle & global \\
      \hline\hline
      $N_c$ & $80345$ & $1\times 10^{10}$ & $1\times 10^{10}$ \\
      \hline
      $a_c$ & $0.35431 \times 10^{-6}$ & 2.1828 & 2.1821 \\
      \hline
      $\gamma_c $ & 3.6432 & 4.5391 & 4.5393 \\
      \hline
      $N_b$ & 19.953 & 19.953 & 41.973 \\
      \hline
      $a_b$ & 6.3031 & 6.3031 & 7.4092 \\
      \hline
      $\gamma_b$ & 1.1773 & 1.1773 & 1.2457 \\
      \hline\hline
      $\chi^2$/d.o.f & 0.4749 & 3.2928 & 2.8030 \\
      \hline
    \end{tabular}
  \end{center}
  \caption{
  Values of fit parameters resulting from the OPAL, Belle and 
  global fits in the GM-VFNS approach together with the value 
  of $\chi^2$ per degree of freedom. 
  }
  \label{tab:fitpars}
\end{table}

\begin{table}[t!]
  \begin{center}
    \begin{tabular}{lllll}
      \hline
      FF set & $B_c(10.52~\mbox{GeV})$ & $B_c(M_Z)$ 
        & $B_b(10.52~\mbox{GeV})$ & $B_b(M_Z)$ \\
      \hline
      OPAL & --- & $4.1739 \times 10^{-2}$ & ---  
        & $8.2474 \times 10^{-2}$ \\
      \hline
      Belle & $6.6476 \times 10^{-2}$ & --- 
        & $8.9244 \times 10^{-2}$ & --- \\
      \hline
      global & $6.6435 \times 10^{-2}$ & $6.4452 \times 10^{-2}$ 
        & $8.3220 \times 10^{-2}$ & $7.7197 \times 10^{-2}$ \\
      \hline
    \end{tabular}
  \end{center}
  \caption{ 
  Values of $c \to \Lambda_c^+$ and $b \to \Lambda_c^+$ 
  fragmentation fractions at $\mu_f = 10.52$~GeV and 
  $\mu_f = M_Z$. 
  }
  \label{tab:fragratios}
\end{table}

\begin{table}[t!]
  \begin{center}
    \begin{tabular}{lllll}
      \hline
      FF set 
        & $x_c(10.52~\mbox{GeV})$ 
        & $x_c(M_Z)$ 
        & $x_b(10.52~\mbox{GeV})$ 
        & $x_b(M_Z)$ \\
      \hline
      OPAL & --- & $0.5389$ & --- & $0.2717$ \\
      \hline
      Belle & $0.5685$ & --- & $0.3063$ & --- \\
      \hline
      global & $0.5685$ & $0.4868$ & $0.3009$ & $0.2666$ \\
      \hline
    \end{tabular}
  \end{center}
  \caption{ 
  Values of average energy fractions for $c \to \Lambda_c^+$ 
  and $b \to \Lambda_c^+$ transitions at $\mu_f = 10.52$~GeV 
  and $\mu_F = M_Z$. 
  }
  \label{tab:eratios}
\end{table}

\begin{figure*}[t!]
\begin{center}
\includegraphics[width=0.48\linewidth]{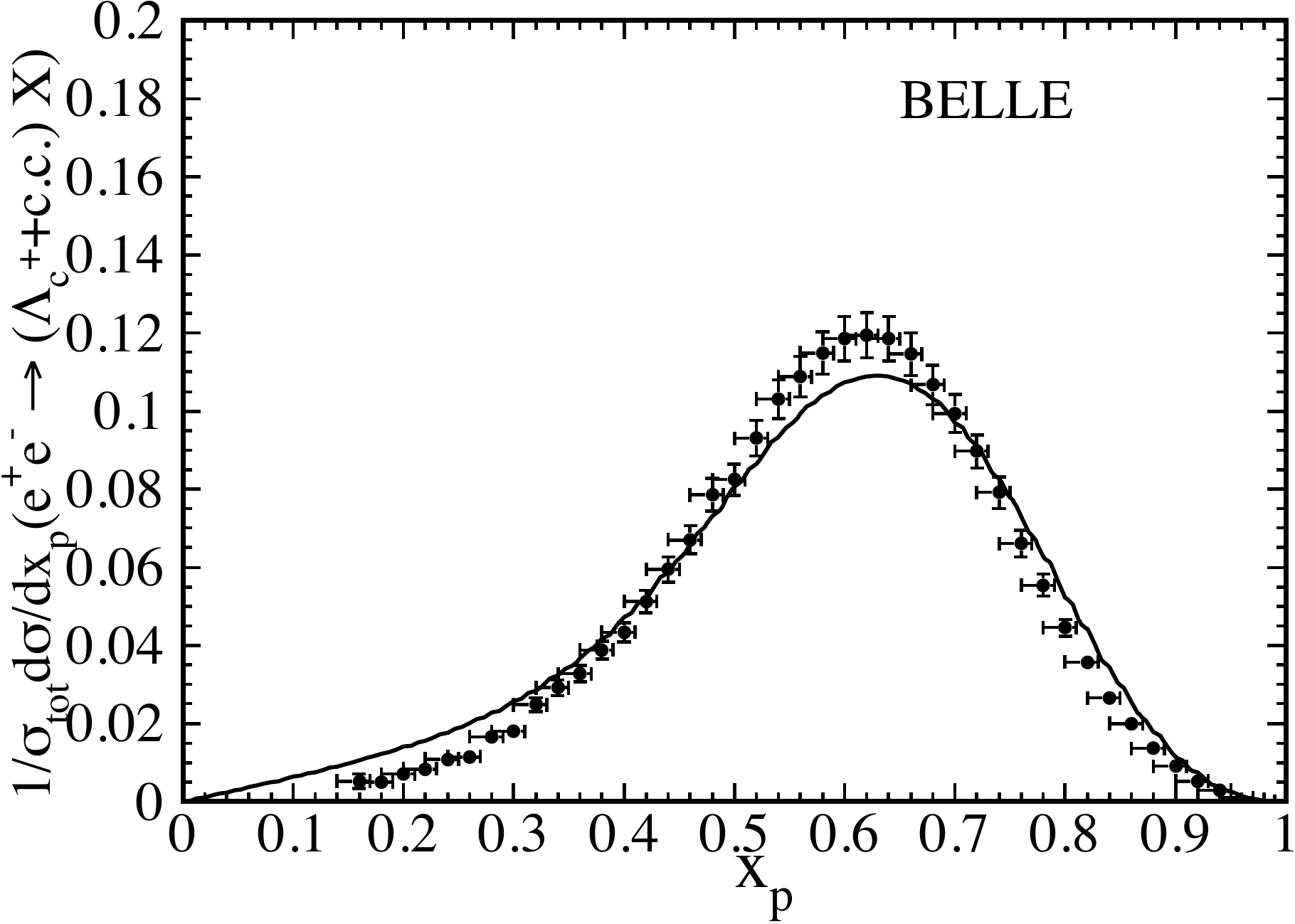}
\end{center}
\caption{
Result of the global fit for the normalized $\Lambda_c^+ + c.c.$ 
production cross section compared with Belle data. 
\label{fig:15} 
}
\end{figure*}

Finally, in our global fit, we use all available data for 
inclusive $\Lambda_c^+ + c.c.$ production in $e^+e^-$ 
annihilation from Belle and OPAL. The resulting values of 
the fit parameters and of $\chi^2$ are included in 
Table~\ref{tab:fitpars}, and the resulting fragmentation  
and average energy fractions are found in 
Tables~\ref{tab:fragratios} and \ref{tab:eratios}, respectively. 
The result of the global fit is compared with OPAL data in 
Fig.~\ref{fig:13} (right). Compared with the OPAL fit, i.e.\ 
without the Belle data, shown in the left part of 
Fig.~\ref{fig:13}, the global fit has a larger $c \to 
\Lambda_c^+$ component. The comparison of the global fit 
with the Belle data is shown in Fig.~\ref{fig:15}. The 
quality of the fit is obviously not perfect. This might be 
connected to the fact that the Bowler ansatz for the 
heavy-quark FFs contains only three free parameters and, 
therefore, is not flexible enough. It is clear that the 
Belle data dominate the fit, since this data set contains 
many more data points with smaller uncertainties. This is 
also reflected by the values of the $c \to \Lambda_c^+$ 
fragmentation fraction shown in Table \ref{tab:fragratios}. 
We, therefore, do not show a separate figure with a comparison 
of the Belle fit with Belle data, since it would be almost 
indistinguishable from Fig.~\ref{fig:15}. 

\begin{figure*}[t!]
\begin{center}
\includegraphics[width=0.48\linewidth]{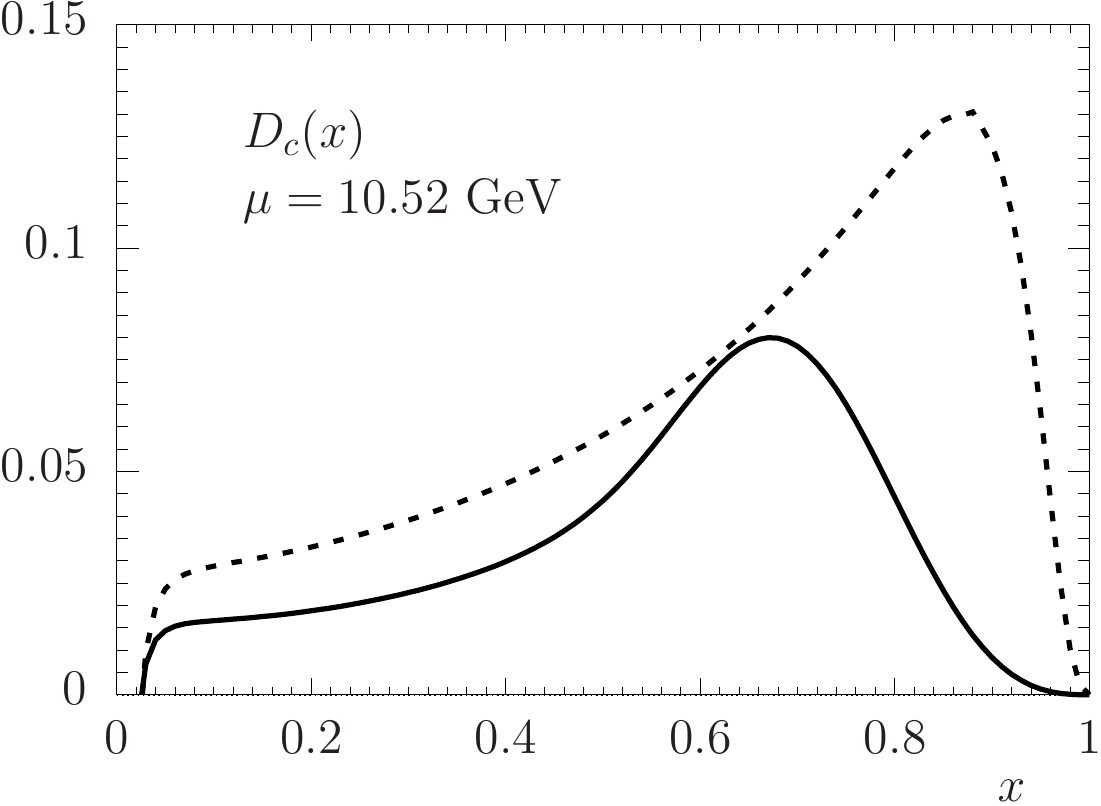}
~~\includegraphics[width=0.48\linewidth]{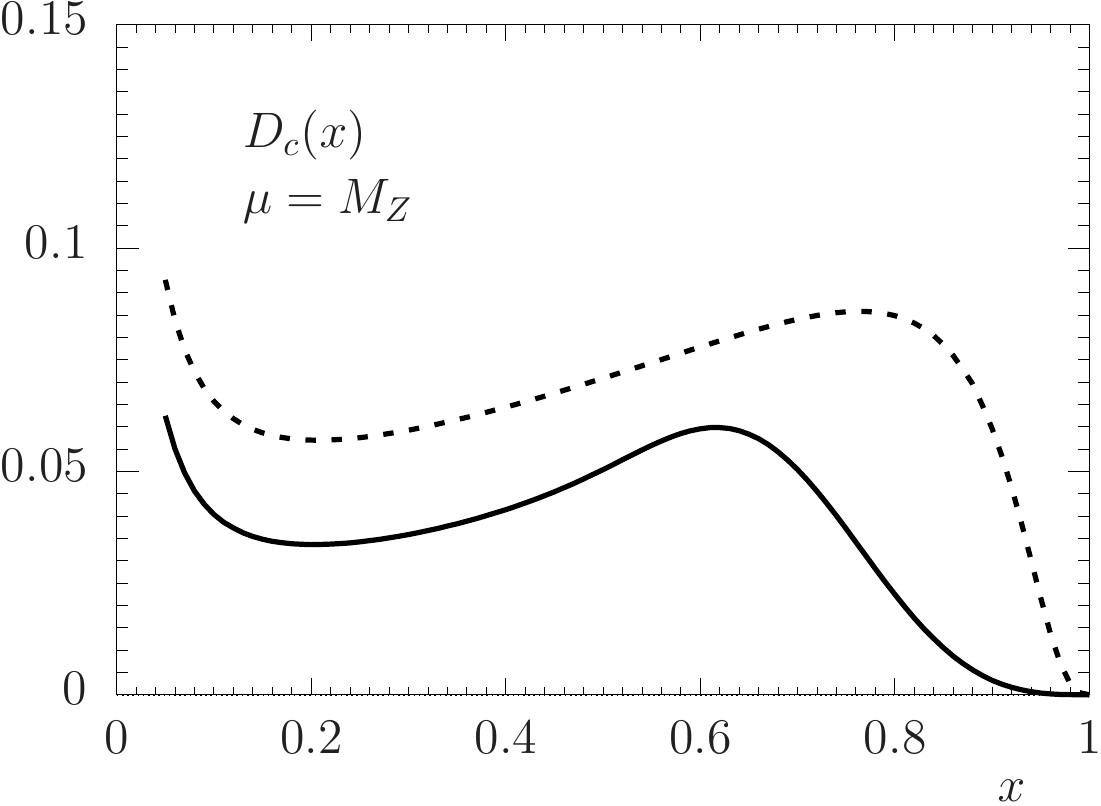}
\end{center}
\caption{
The fragmentation function for $c \to \Lambda_c^+$ at 
scales $\mu=10.52$~GeV (left) and $\mu=M_Z$ (right). The 
full curves show the new fit described in this work, the 
dashed lines represent the fit of Ref.~\cite{Kniehl:2006mw}.
\label{fig:14} 
}
\end{figure*}

\begin{figure*}[t!]
\begin{center}
\includegraphics[width=0.48\linewidth]{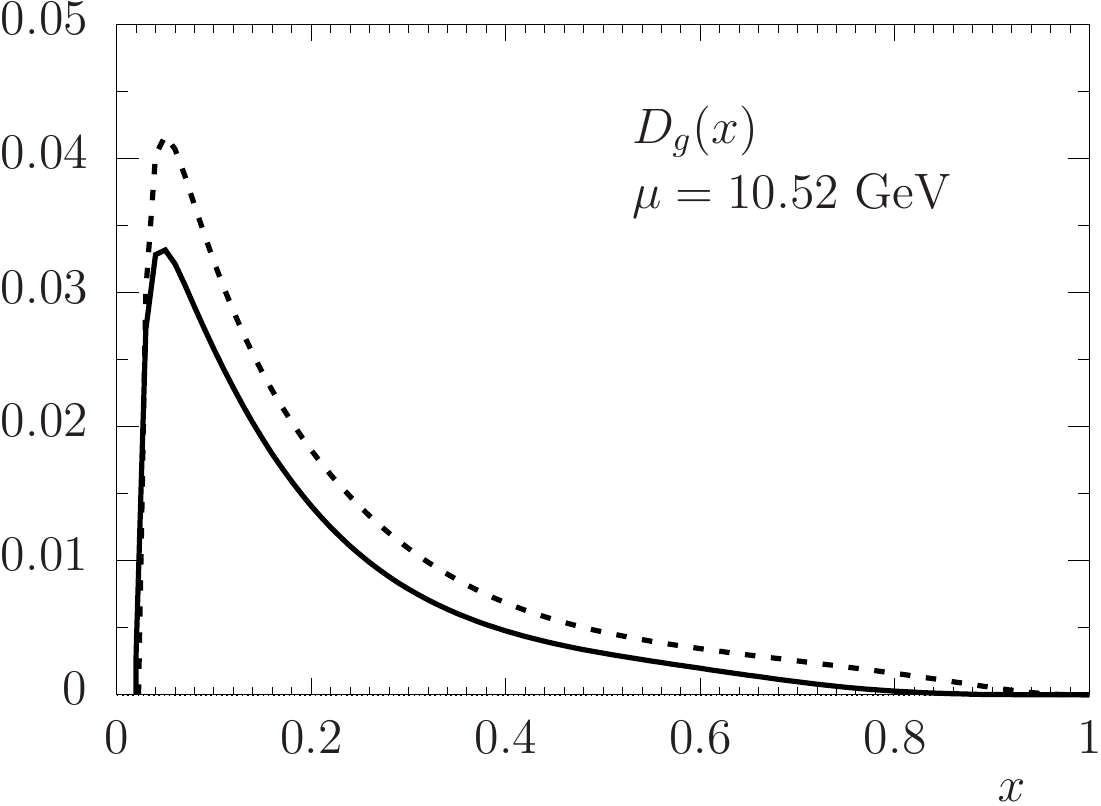}
~~\includegraphics[width=0.48\linewidth]{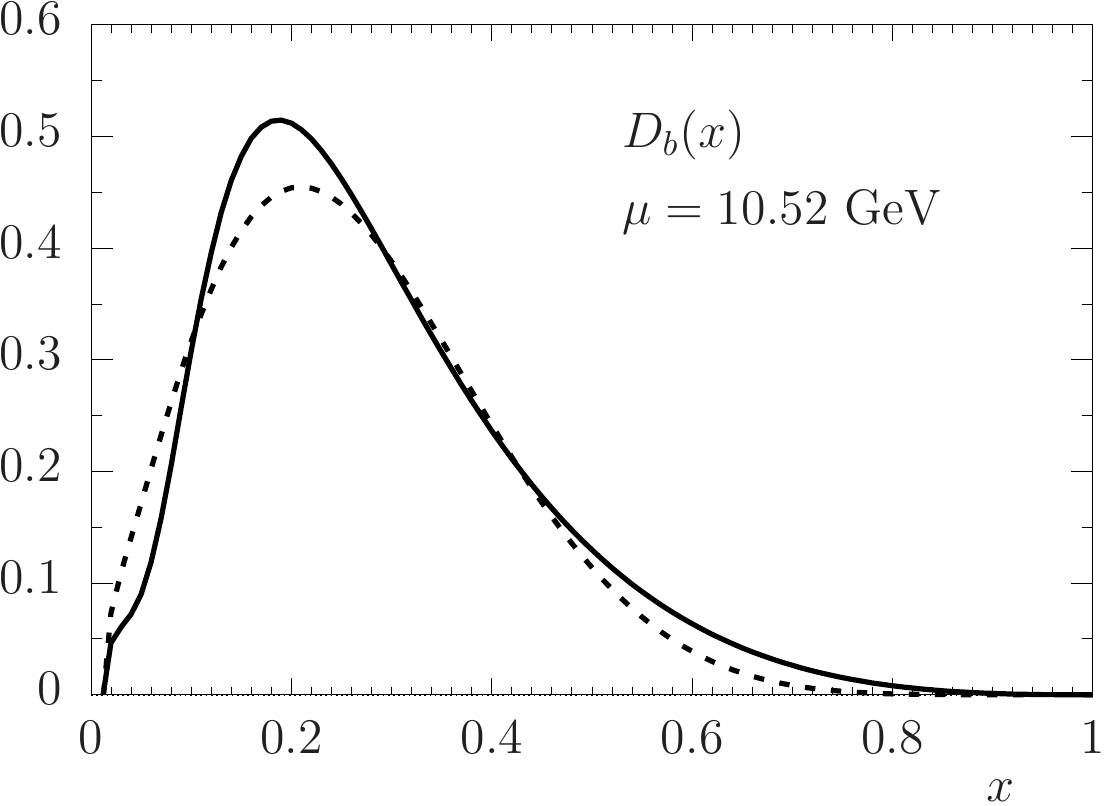}
\end{center}
\caption{
The fragmentation function for $g \to \Lambda_c^+$ (left) and 
$b \to \Lambda_c^+$ (right) at $\mu=10.52$~GeV. The full curves 
show the new fit described in this work, the dashed lines 
represent the fit of Ref.~\cite{Kniehl:2006mw}. 
\label{fig:16} 
}
\end{figure*}

A direct comparison of the old and new FFs is shown in 
Figs.~\ref{fig:14} and \ref{fig:16}. We note that both 
shape and normalization of $D_c(x)$ have changed. The 
maxima of the new FFs appear at smaller values of $x$. The 
normalization is reduced, since we have rescaled the OPAL 
data by the factor 0.6929 to match the new branching ratio 
of the $\Lambda_c^+ \to \pi^+K^-p$ decay. The differences 
between the two fits for the $D_g$ and $D_b$ FFs 
(Fig.~\ref{fig:16}) are smaller. We note that the 
$g \rightarrow \Lambda_c^+$ FF has its maximum at rather 
low $x$ and does therefore not play an important role for 
the comparison with the available $e^+e^-$ data which, in 
turn, means that $D_g(x)$ is not well constrained by these 
data. One would need to perform fits to $pp$ data in order 
to improve our knowledge of the gluon FF. 
 
\section{$\Lambda_c^{\pm}$ production wit new FF 
and comparison with LHC data}

The FFs obtained in the fits described in the previous 
section have been converted into a grid, which subsequently 
is used for the calculation of inclusive $\Lambda_c^{\pm}$ 
production cross sections in $pp$ collisions at the LHC to 
be compared with data from LHCb, ALICE and CMS measurements. 
Corresponding results are presented now. 

\begin{figure*}[b!]
\begin{center}
\includegraphics[height=0.43\textheight]{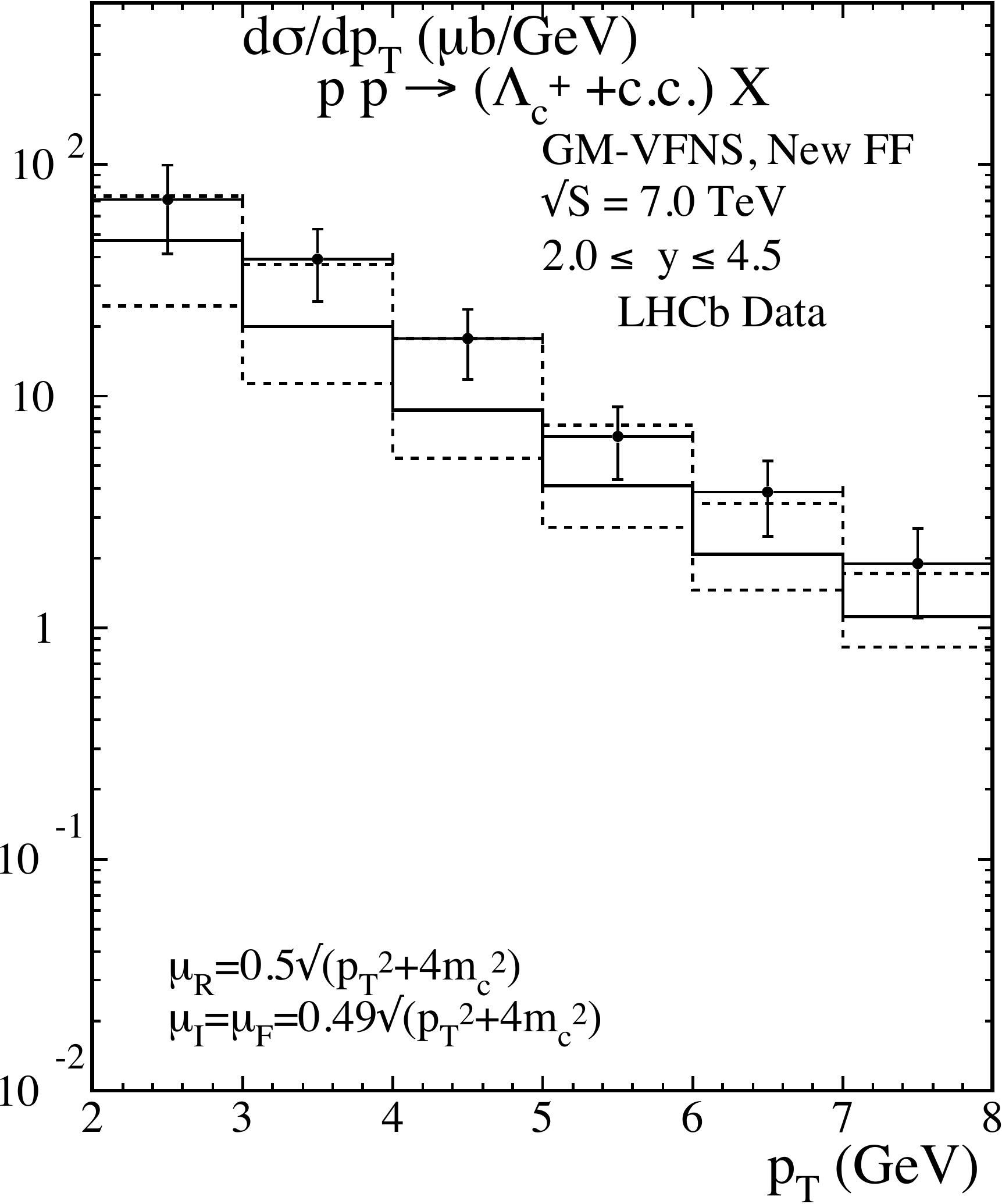}
\includegraphics[height=0.43\textheight]{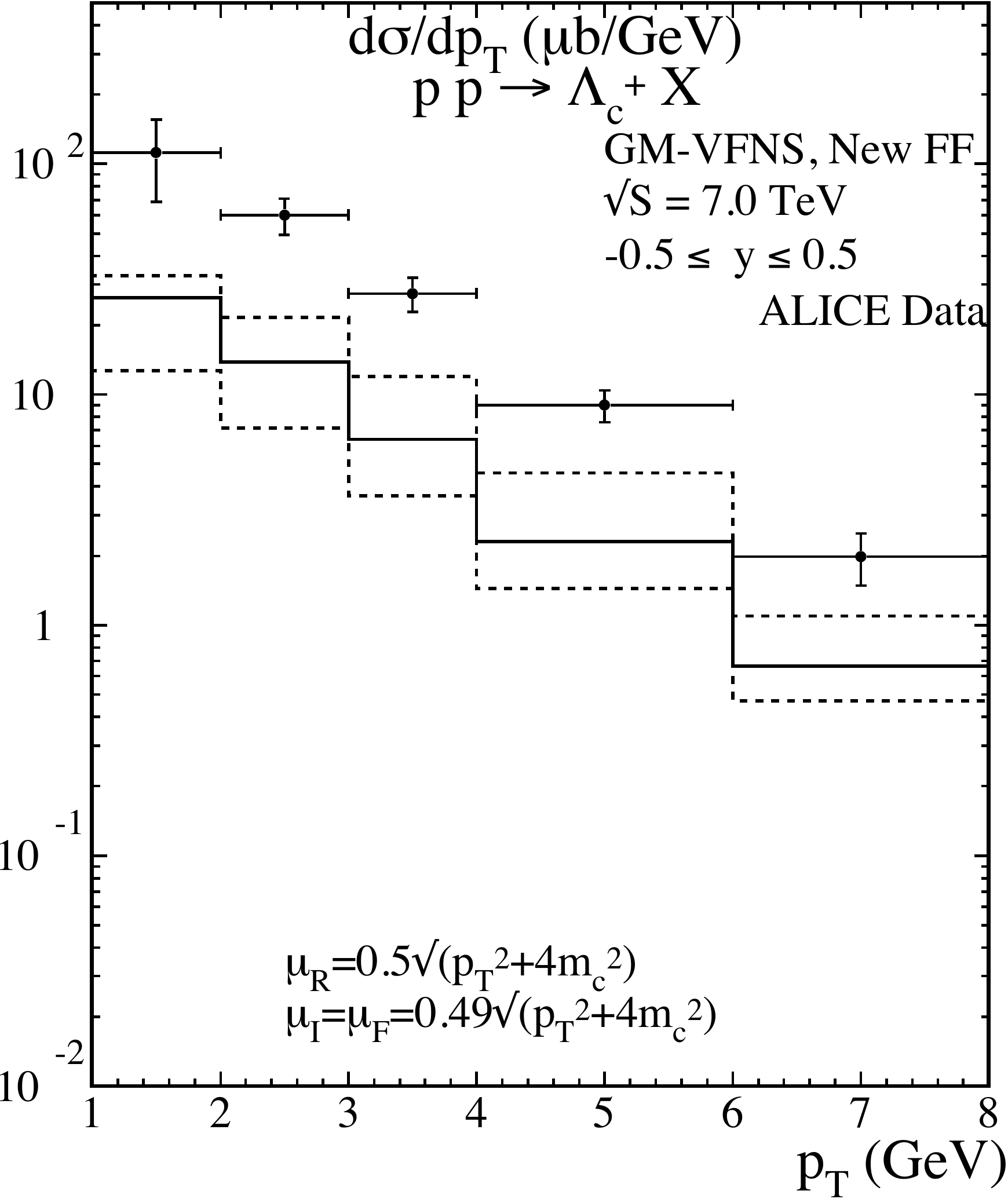}
\\
\includegraphics[height=0.43\textheight]{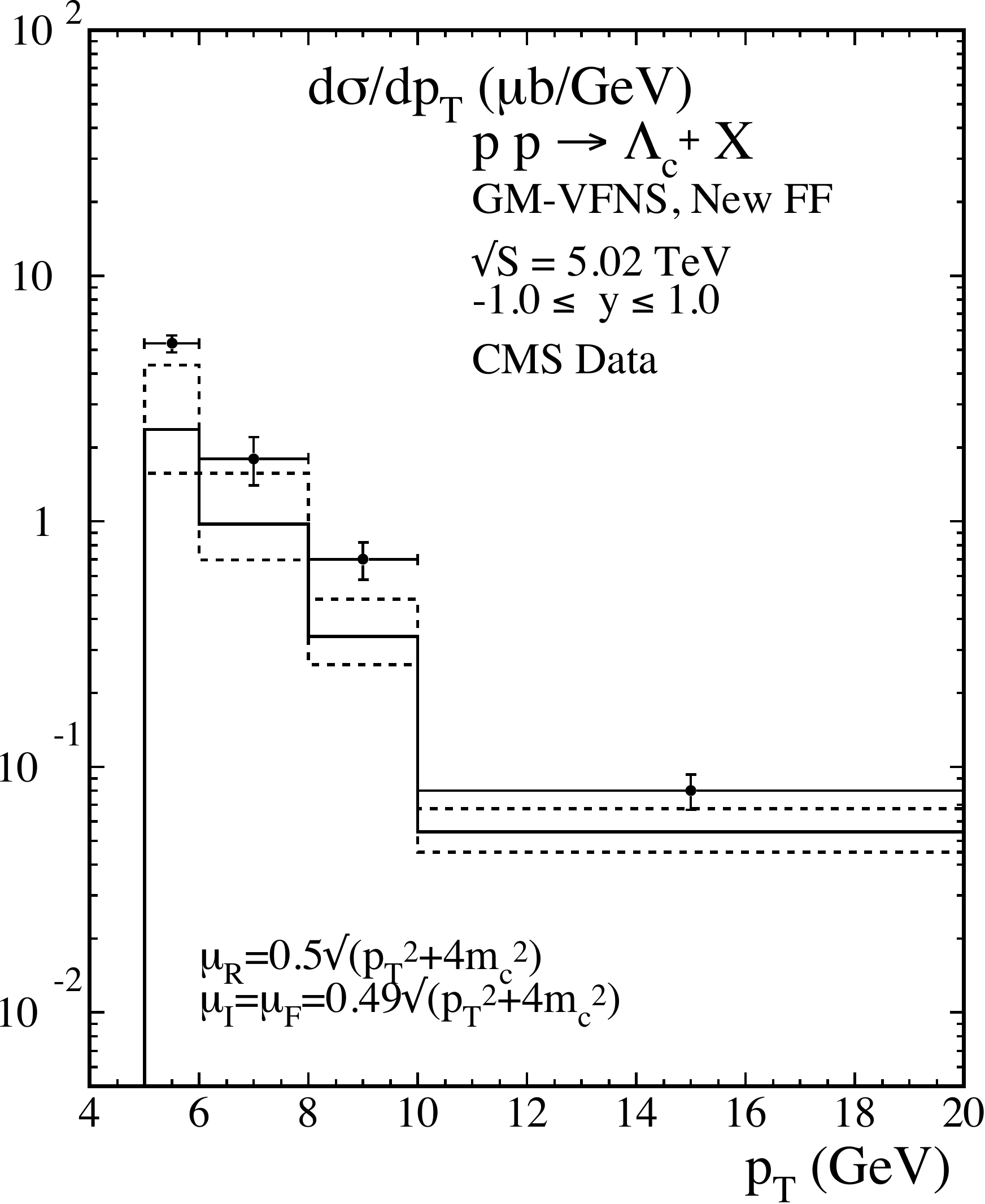}
\end{center}
\caption{
Differential $\Lambda_c^{\pm}$ production cross sections 
evaluated with the new FF fit compared with data from 
LHCb (upper left), ALICE (upper right) and CMS (lower). 
\label{fig:10} 
}
\end{figure*}

In Fig.~\ref{fig:10}, our results are compared with data 
from the LHCb, ALICE and CMS collaborations in the same way 
as this was done in Figs.~\ref{fig:1}, \ref{fig:4} and 
\ref{fig:7} (left sides). The predicted cross sections for 
the three experiments come out similar to the earlier results 
in the previous section, but are slightly smaller. For example, 
the LHCb cross section in Fig.~\ref{fig:10} is smaller than the 
one in Fig.~\ref{fig:1} by approximately $15~\%$ in the first 
$p_T$ bin and $35~\%$ in the last $p_T$ bin. The reduction of 
the cross section $d\sigma/dp_T$ is similar for ALICE and CMS 
(see upper right and lower panels of Fig.~\ref{fig:10}). This 
means that also the $\Lambda_c^+/D^0$ ratio is reduced by 
about the same amount when using the new FFs. This change 
is not very large, and the comparison of theory predictions 
with data is qualitatively the same with the new FF. For the 
LHCb data (Fig.~\ref{fig:10}, upper left panel) and for the 
CMS data (Fig.~\ref{fig:10}, lower panel) the measured cross 
sections $d\sigma/dp_T$ for $\Lambda_c^{\pm}$ production are still 
very close to the upper border of the theory uncertainty band. 
Only the ALICE data (Fig.~\ref{fig:10}, upper right panel) 
are significantly higher, by factors of 2 to 4, than the 
cross sections predicted from theory. It is very unlikely 
that this large discrepancy can be explained in the usual 
framework by realistic FFs for the $\Lambda_c^+$ without 
destroying agreement with other data. This suggests that 
the ALICE data violate predictions based on the universality 
of the fragmentation process by a large factor, whereas the 
data from LHCb \cite{Aaij:2013mga} and CMS \cite{Sirunyan:2019fnc} 
appear to be compatible with universality.

\section{Discussion and summary}

In Ref.~\cite{He:2019tik}, an attempt was made to explain the 
large $\Lambda_c^+/D^0$ ratio seen in the ALICE measurements 
as arising from processes where $\Lambda_c^+$ is produced from 
the decay of excited charm baryon states, which are not seen 
in the Belle data of $e^+e^-$ annihilation. Actually, in the 
Belle data, only about $50\%$ of the $\Lambda_c^{\pm}$ production 
rate originates from direct $c \to \Lambda_c^+$ fragmentation;  
the other half is due to decays of excited charm baryon states. 
However, it is difficult to understand that this fraction 
should be larger in $pp$ collisions by a large factor, or 
that the production strength of the higher resonances discovered 
already in the Belle experiment should be so much larger in 
the ALICE $pp$ collision experiment. This is an experimental 
question which could be answered only in the respective 
experiments.

Another possibility to enhance inclusive $\Lambda_c^{\pm}$ 
production in $pp$ collisions is suggested by models that 
include colour reconnection mechanisms \cite{Christiansen:2015yqa} 
in the usual PYTHIA hadronization scheme. This implies that 
the final partons in the string fragmentation are considered 
to be colour-connected in such a way that the total string 
length becomes as short as possible. In 
Ref.~\cite{Sirunyan:2019fnc}, it is shown that 
this model is consistent with the CMS result for the 
$\Lambda_c^+/D^0$ ratio, but it is not sufficient to explain 
the strong enhancement of this ratio in the ALICE data 
\cite{Acharya:2017kfy}. 

It seems unlikely that higher-order corrections, which go 
beyond the available NLO calculations, could help to obtain 
a better agreement of data with theory. NNLO fits for 
$D$-meson FFs have been studied recently in 
Refs.~\cite{Shoeibi:2017zha,Salajegheh:2019nea}. However, 
such corrections are expected to affect all measurements 
in a similar way. Also alternative approaches like 
$k_t$-factorization \cite{Maciula:2018iuh}, are not able 
to explain the observed discrepancies. The different coverage 
of kinematic variables, $p_T$ and $y$, is not very different 
in the available data sets. On the experimental side, however, 
more data differential both in $p_T$ and $y$ would be extremely 
helpful, at least to clarify that data from different experiments 
are compatible with each other in the kinematic regions where 
they overlap.


\section*{Acknowledgment}

We thank M.\ Niiyama for providing us with data of the 
Belle experiment in numerical form. I.S.\ would like to 
acknowledge the Mainz Institute of Theoretical Physics 
(MITP) for its hospitality and support. The work of B.A.K.\ 
was supported in part by the German Federal Ministry for 
Education and Research BMBF through Grant No.\ 05H18GUCC1 
and by the German Research Foundation DFG through Research 
Unit FOR~2926 ``Next Generation Perturbative QCD for Hadron 
Structure: Preparing for the Electron-Ion Collider'' with 
Grant No.~KN~365/14-1.



\end{document}